\documentclass[a4paper,11pt]{article}
\usepackage{amsmath}
\usepackage{jheppub} 
\usepackage{multicol}
\usepackage{scrextend}
\usepackage[dvipsnames]{xcolor}
\usepackage{tikz}
\usepackage[T1]{fontenc} 

\newcommand{\bZ}{\mathbb{Z}}

\newcommand{\cO}{\mathcal{O}}
\newcommand{\cT}{\mathcal{T}}

\newcommand{\hoo}{h^{11}}
\newcommand{\hoomin}{h^{11}_{\text{min}}}
\newcommand{\hoomax}{h^{11}_{\text{max}}}
\newcommand{\sdoc}{S_{\Delta_1^\circ}}
\newcommand{\sdtc}{S_{\Delta_2^\circ}}
\newcommand{\sdc}{S_{\Delta^\circ}}
\newcommand{\dc}{{\Delta^\circ}}
\newcommand{\doc}{{\Delta_1^\circ}}
\newcommand{\dtc}{{\Delta_2^\circ}}

\def\ge{E}
\def\gso{SO}
\def\gsu{SU}
\def\gsp{Sp}
\def\gf{F}
\def\gg{G}

\title{\boldmath Machine Learning in the String Landscape}

\author[a]{Jonathan Carifio,}
\author[a]{James Halverson,}
\author[a,b,c]{Dmitri Krioukov,}
\author[a]{ and Brent D. Nelson}

\affiliation[a]{Department of Physics, Northeastern University,\\Boston, MA 02115, USA}
\affiliation[b]{Department of Mathematics, Northeastern University,\\Boston, MA 02115, USA}
\affiliation[c]{Department of Electrical and Computer Engineering, Northeastern University,\\Boston, MA 02115, USA}

\abstract{

  \vspace{.2cm}
  \noindent We utilize machine learning to study the string landscape.
  Deep data dives and conjecture generation are proposed as useful
  frameworks for utilizing machine learning in the landscape, and
  examples of each are presented.  A decision tree accurately predicts
  the number of weak Fano toric threefolds arising from reflexive
  polytopes, each of which determines a smooth F-theory
  compactification, and linear regression generates a previously
  proven conjecture for the gauge group rank in an ensemble of
  $\frac43 \times 2.96 \times 10^{755}$ F-theory compactifications.
  Logistic regression generates a new conjecture for when $E_6$ arises
  in the large ensemble of F-theory compactifications, which is then
  rigorously proven. This result may be relevant for the appearance
  of visible sectors in the ensemble. Through conjecture generation, machine learning
  is useful not only for numerics, but also for rigorous results.

  \vspace{1cm}
}

\begin{document} 
\maketitle
\flushbottom

\section{Introduction}

String theory is perhaps the most promising candidate for a unified
theory of physics. As a quantum theory of gravity that naturally gives
rise to general relativity at long distances and the building blocks
for realistic particle and cosmological sectors, it satisfies a number
of non-trivial necessary conditions for any unified theory. In fact,
it is the only known theory that satisfies these necessary
conditions. However, its extra dimensions of space allow for many
compactifications to four dimensions, which give rise to a large
landscape of vacua that may realize many different incarnations of
particle physics and cosmology. Taming the landscape is therefore a
central problem in theoretical physics, and is critical to making
progress in understanding unification in string theory.

In this paper, we treat the landscape as what it clearly is: a big
data problem. In fact, the data that arise in string theory may be the
largest in science. For example, in type IIb compactifications there
are many possible Ramond-Ramond background fluxes that depend on the
topology of the extra dimensions. This is the context for the
oft-quoted $O(10^{500})$ type IIb flux vacua
\cite{Bousso:2000xa,Ashok:2003gk,Denef:2004ze}, though in recent years
this number has grown to an estimated $O(10^{272,000})$
\cite{Taylor:2015xtz} . Recently \cite{Halverson:2017ffz}, the number
of geometries has grown beyond the early flux number to
$\frac43\times 2.96 \times 10^{755}$, which is known to only be a
lower bound.  Dealing with these large numbers is exacerbated by the
computational complexity of the landscape
\cite{Denef:2006ad,Cvetic:2010ky,Denef:2017cxt,Bao:2017thx}, making it
even more clear that sophisticated techniques are required.

How should one treat ensembles so large that they forbid explicit
construction?  One analytic technique is algorithmic
universality. Rather than deriving universality from exploration of a
constructed ensemble, algorithmic universality is derived instead from
a concrete construction algorithm. This idea, while obvious, was
exemplified in \cite{Halverson:2017ffz} and used to demonstrate universal features in the
ensemble of $10^{755}$ F-theory geometries. For example, knowledge of
the construction algorithm demonstrated that the probability $P_{NHC}$
of having clusters of non-trivial seven-branes is
$1 > P_{NHC} \geq 1-1.07\times 10^{-755}$, and also that the probability
$P_G$ of having a particular minimal geometric gauge group $G$ with
$rk(G)\geq 160$ is $1 > P_G \geq .999995$. This degree of control over
such a large ensemble is ideal, but in many physically interesting cases
such precise construction algorithms are not yet known. In this case,
one must either develop such construction algorithms, which may not
always be possible, or utilize other techniques.

The other possibility is to use numerical techniques from data
science, and in some cases we will see that these numerical techniques
can lead to rigorous results. Specifically, we will employ
modern machine learning techniques to study the landscape.  Machine
learning is a broad term for a wide variety of techniques that allow a
computer to develop prediction models for complex datasets or to
extract features. It has led to veritable revolutions in a number of
fields, from genotyping and gene expression 
to oceanography and climate science. We will provide a review of basic machine learning techniques in Section~\ref{sec:mlforst}.

It is easy to imagine a number of broad ways in which machine learning
could be of use in string theory, as well as mathematics and
theoretical physics more broadly.
\begin{itemize}
\item \textbf{Deep Data Dives.} Via training a model on a subset of an
  ensemble, it is sometimes feasible to make high accuracy feature
  predictions that are much faster than conventional techniques,
  allowing for far greater exploration of the dataset.
\item \textbf{Conjecture Generation.} The decision function of a
  trained model may naturally lead to a sharp conjecture that can be
  rigorously proven.
\item \textbf{Feature Extraction.} When input data is of high dimensionality, or exhibits redundant information, models can identify those properties ({\em features}) of the data that are most correlated with desired outcomes. This is often one of the primary goals of landscape surveys in string theory.
\end{itemize}
Many other possibilities may arise as new machine learning techniques
are developed by computer scientists.

In this paper we will present one example of a data dive, and two of
conjecture generation. The data dive and one of the generated
conjectures are known results, while the other generated conjecture is
a genuinely new result that would have been more difficult to obtain
without machine learning.

\vspace{.5cm}
\noindent \textbf{Summary of Results.}

The layout of our paper and the summary of our results are as follows.

The data dive is used to study the geometries relevant for certain
four dimensional F-theory compactifications.  Specifically, in Section~\ref{sec:3dPoly}
we use machine learning to compute the number of fine regular star
triangulations (FRST) of three-dimensional reflexive polytopes, each
of which determines a smooth F-theory compactification to
four-dimensions, amongst other possible applications. These results
compare well to known results about the number of FRST of those
polytopes.  In~\cite{Upcoming}, similar techniques will be used in the
case of four-dimensional reflexive polytopes, which will provide
an upper-bound estimate of the number of Calabi-Yau threefolds in the Kreuzer-Skarke set.

The conjecture generation arises in the context of the ensemble of
$\frac43 \times 2.96 \times 10^{755}$ F-theory geometries. We utilize
random sampling to generate data that will be used to train models. In
Section \ref{sec:rk} we train models that accurately predict the rank
of the gauge group in each geometry. Study of the decision function
and properties of the data structure give rise to a sharp conjecture
that may be rigorously proven; in fact, the result was known, though
the conjecture was not previously arrived at via machine learning. The
general 
process by which that conjecture was generated is applicable in
other contexts.  In Section~\ref{sec:e6} we use this process to generate a
conjecture regarding conditions under which $E_6$ arises in the
ensemble. The conjecture is proven, which leads to a computation of
the probability of $E_6$ in the ensemble. Comparing to five~independent
sets of $2,000,000$ random samples, the predictions match quite well.

We find it promising that new rigorous results may arise from
conjectures that would be difficult to arrive at without the help of
machine learning. Further areas of promise for utilizing machine learning in string theory will be suggested in the concluding section.

\vspace{1cm} While we were completing this work, \cite{He:2017aed} and
\cite{Ruehle:2017mzq} appeared. Those works also utilize learning to
study the landscape, but they differ in specific techniques and
physical applications.  \cite{Krefl:2017yox} used machine learning to
find the minimum volumes of Sasaki-Einstein manifolds.

\section{Machine Learning for Theoretical Physicists\label{sec:mlforst}}

Since machine learning is not part of the everyday lexicon of a theoretical physicist,
in this section we would like to review the basics of the subject, including all of
the techniques that we utilized in this paper. The subject has a rich literature; for
a more in-depth introduction we recommend the textbooks \cite{MLBook2,MLBook1}. Here, we focus on explaining the basic ideas behind supervised learning, unsupervised learning, and model evaluation.

\subsection{Supervised Learning}

Machine learning is a set of algorithms that trains on a data set in order to make predictions on unseen data. As such, simple least-squares regression -- a tool with which every scientist is familiar -- is seen as the most humble form of machine learning. Modern techniques can be incredibly sophisticated, but it is worth remembering that the output from most machine learning algorithms is a function (commonly called a {\em model}): it takes a specified set of inputs and produces a unique output value for the characteristic in question.

The procedure described above is commonly referred to as {\em supervised} machine learning, or simply {\em supervised learning}. For supervised learning, the training step is performed by allowing the algorithm to experience a number of input $\to$ output pairs. This is the most common (and generally most successful) form of machine learning. All of the results in this paper, as well as those that have appeared recently in the literature \cite{He:2017aed,Krefl:2017yox,Ruehle:2017mzq}, arise from the application of supervised machine learning.

For physics applications, the input data may be mathematical and abstract. Examples may include the vertices of a polytope in $\mathbb{Z}^3$ or $\mathbb{Z}^4$ or the tensor of intersection numbers of two-cycles on a Calabi-Yau threefold. The outputs are generally meaningful physical quantities, such as the presence of an orientifold involution or the rank of a gauge group.

In order to effectively utilize supervised machine learning, the physicist must have identified these physically relevant output variables so as to prepare a relevant training sample in the first place. When no such outputs are identified, we refer to the procedure as {\em unsupervised machine learning}.

\subsection{Unsupervised Learning}

As no particular output is specified in unsupervised learning, the goal of an algorithm is often more modest. For example, one might employ a {\em clustering algorithm} that seeks to group inputs according to which cases are most `similar' in some fashion. Another common goal of unsupervised learning is {\em dimensionality reduction}, in which a large dataset of high dimensionality is transformed into a smaller dataset through a form of coarse-graining, with a goal of retaining the essential global properties of the data. 
Both clustering and dimensionality reduction are employed in supervised learning as well, where the latter is often referred to as {\em factor analysis}, or {\em principal component analysis}.

The datasets which concern us in string theory are often cases in which the data is itself constructed through well-defined mathematical algorithms, with a specific physics goal in mind. As such, there is less need for unsupervised learning in as much as the outcome of unsupervised machine learning is often merely a step in the direction of more effective supervised learning. Again, the lack of a specific goal, or output variable, in unsupervised learning makes it difficult to answer the question of how well the algorithm has performed. This is called {\em model evaluation}.

\subsection{Model Evaluation}

In supervised learning, model evaluation is more straight-forward. An obvious manner in which to proceed is as follows. Let us take the total set of cases in which input $\to$ output pairs are known, and divide it into a {\em training set} and a {\em test set}. This is called a {\em train-test split}. The model is then developed through training on the training set. Once created, the model is then evaluated on the test set. The performance of the model can then be evaluated through normal statistical means.

When computational cost is not at issue, a {\em $k$-fold cross-validation} procedure is a better training method. In this case, the data is divided into $k$ equal subsets. Our model will now be trained $k$ separate times, in each reserving only one of the $k$ partitions (or `folds') for testing, and using the other $k-1$ folds for training. This method has several advantages. First, it minimizes training-sample bias, in which our training data happens to be different qualitatively from our testing data. Second, while a simple train-test split might train on just 50\% of the data, a ten-fold cross-validation (which we use in this paper) will train on 90\% of the available data. Finally, we will evaluate the efficacy of our model multiple times on different independent pieces of our data, thereby giving us a measure of how robust our model is to perturbations on the data.

The correct model evaluation metric depends on the type of algorithm being employed. If the goal is one of clustering, such as the {\em binary classification} problem, we are usually concerned with the {\em accuracy} of the model. Thus, if the goal is to predict, given the input data, whether or not a certain number is non-vanishing, then the model will return a function which varies from zero to one. The accuracy can then be computed from the fraction of ``true'' cases in the test data for which the model returns unity. A more generalized evaluation tool is the {\em confusion matrix}, which also returns the number of ``false-positives'' and ``false-negatives''. 

In physics applications, we are more often trying to predict a continuous real number, which is a form of {\em regression analysis}. The evaluation of regression algorithms is typically a statistical goodness-of-fit variable. We will discuss a number of such algorithms below. 

\subsection*{Techniques Used in This Paper}
There are a number of supervised learning algorithms that we use in this paper: Linear Regression, Logistic Regression, Linear Discriminant Analysis, k-Nearest Neighbors, Classification and Regression Tree, Naive Bayes, and Support Vector Machine. In the following we will briefly
describe the workings of each algorithm, as well as the pros and cons of each, in cases that
they are known.

\begin{description}
\item[Linear Regression (LIR)] is the analog of least-squares regression for the case of multiple input variables. If the output quantity is denoted by $y$, then the model seeks a set of $w_i$ and an intercept $b$ such that $y = b+ \sum_i^n w_i x_i$, where $n$ is the number of input properties, labeled by $x_i$. This method can be generalized to allow for constraints on the allowed weights -- perhaps to reflect known physical requirements.
\item[Logistic Regression (LR)] gets its name from its reliance on the {\em logistic function}, sometimes also referred to as the {\em sigmoid function}. As this function interpolates between zero and unity, it is often used by statisticians to represent a probability function. As such, logistic regression (despite its name) is most often used in binary classification problems. While a linear method, logistic regression is more general than linear regression in that predictions are no longer merely linear combinations of inputs. The advantages of this technique include its relative ease of training and its suitability for performing factor analysis. In addition, logistic regression is more resistant to the peril of {\em overfitting} the input data, in which noise (random fluctuations) in the input data is incorporated into the model itself, thereby limiting its ability to correctly predict on newly encountered data. 
\item[k-Nearest Neighbors (KNN)] is an algorithm that develops a prediction for a particular output by looking at the outputs for the $k$ closest `neighbors' in the input space. Once a metric on the input parameter space is specified, the $k$-closest neighbors are identified. The predicted output may then be as simple as the weighted average of the values of the neighbors. This method has the advantage of extreme conceptual simplicity, though it can be computationally costly in practice. The method is thus best suited to input data with low dimensionality -- or where factor analysis reveals a low effective dimensionality. The method will be less useful in cases where the data is of high dimensionality, where some data spans a large range of absolute scales, or where the data cannot be readily expressed in the form of a simple $n$-tuple of numbers.
\item[Classification and Regression Tree (CART)] is the name given to {\em decision tree} approaches to either classification or regression problems. These algorithms divide up the input feature space into rectangular regions, then coarse grain the data in these regions so as to produce ``if-then-else'' decision rules. In so doing, such methods generally identify feature importance as an immediate by-product of the approach. Decision trees are not linear in the parameters: they are able to handle a mix of continuous and discrete features and the algorithms are invariant to scaling of the data. By their very nature, however, they are ill-suited to extrapolation to features beyond those of the training data itself.
\item[Naive Bayes (NB),] as the name suggests, seeks to compute the probability of a hypothesis given some prior knowledge of the data. More specifically, we are interested in the most important features of the data, and we assume that these features are conditionally independent of one another (hence `Naive' Bayes). A great number of hypothesis functions are created from the data features, and the one with the maximum {\em a posteriori} probability is selected. This linear method is generally used in classification problems, but can be extended to continuous prediction in {\em Gaussian Naive Bayes}. Here, each feature is assigned a distribution in the data; for example, a Gaussian distribution, which can be classified simply by the mean and the variance of that feature across the dataset. This makes Gaussian NB very efficient to implement on data sets in which the input data is effectively continuous in nature, or of high dimensionality. However, if input data is known to be highly correlated, we should expect NB algorithms to perform poorly relative to other methods.
\item[Linear Discriminant Analysis (LDA)] has similarities to both logistic regression and Gaussian naive Bayes approaches. It assumes that the data has a Gaussian distribution in each feature, and that the variance for each feature is roughly equivalent. Like NB methods, it then builds and evaluates a probability function for class membership, and is thus generally used in classification problems. Despite the name, LDA approaches can be effectively non-linear, by constructing a {\em feature map} that transforms the raw input data into a higher-dimensional space associated with {\em feature vectors} (sometimes called the {\em kernel trick}). In this, LDA algorithms share many methods with principal component analysis (PCA). The latter is often described as `unsupervised' in the sense that its goal is to find directions in feature space that maximize variance in the dataset, while ignoring class labels, while LDA is `supervised' in the sense that it seeks to find directions in feature space that maximize the distance between the classes.
\item[Support Vector Machine (SVM)] is another technique that involves feature space, and is often one of the more powerful supervised learning algorithms. The feature space is partitioned using linear separation hyperplanes in the space of input variables. The smallest perpendicular distance from this hyperplane to a training data point is called the {\em margin}. The optimal dividing hyperplane is thus the one with the maximal margin. The training data that lie upon these margins are known as the {\em support vectors} -- once these points are identified, the remainder of the training data becomes irrelevant, thereby reducing the memory cost of the procedure. At this point we have a support vector classifier (SVC). The method can be generalized by using the kernel method described above, which is necessary in cases for which the desired outputs overlap in feature space. In such cases the generalization is referred to as the support vector machine algorithm. SVM can be adapted to perform regression tasks as well (support vector regression).
\end{description}

\subsection*{Techniques Not Used in This Paper}

In the course of preparing this work, other examples of using machine learning techniques in string theory have appeared. The techniques in these cases generally involve the use of {\em neural networks}, whose properties we briefly describe in this subsection. 

The basic building block of a neural network is the {\em perceptron}, which can be described as a single binary classification algorithm. These algorithms are generally linear, and thus each of the techniques described above, and used in this paper, can be thought of as individual perceptrons. Multi-layer perceptron (MLP) models are generalization of the linear models described above in that multiple techniques are layered between the input data and the output predictions. It is these collections of perceptrons that tend to be called neural networks.
The simplest cases are feed-forward neural networks, in that each layer delivers its output as the input to the subsequent layer. More sophisticated approaches allow flow of information in both directions, thereby producing feedback. These latter cases more closely approximate the structure of true biological neural networks.

Like the cases described above, neural networks themselves can be supervised or unsupervised. That is, neural networks can be trained on data in which input $\to$ output pairs are known, or allowed to perform clustering or dimensionality reduction tasks on data where no outputs are specified. Furthermore, the individual layers in the MLP can be chosen by the model-developer in advance, or allowed to evolve through feedback, with the latter case introducing an element of unsupervised learning into the MLP.\footnote{It is common to use the phrase ``deep learning'' to indicate any machine learning technique that involves a neural network model, though often this phrase is restricted to those cases which involve some element of unsupervised learning.}

Neural networks and single-model methods generally fare comparably, though each has its place and relative advantages. By adding layers and feedback channels, a neural network can be designed with a large number of free, tunable parameters. This can lead to the problem of overfitting, however. On the other hand, such frameworks are generally suited to many forms of input data, or highly heterogeneous input data. A key disadvantage is the fact that the output of a neural network training tends to be a `black box', which makes such techniques less useful for feature extraction or conjecture generation, though still quite powerful for deep data dives.\footnote{These same criticisms could be leveled against the most sophisticated SVM techniques, which are often comparable to neural networks in complexity.} For these various reasons we have chosen to work with the single-algorithm approach, and will generally use the simplest such approaches to achieve the goals of this paper.

\section{Data Dive: The Number of Smooth F-theory Compactifications\label{sec:3dPoly}}
In this section we use machine learning to estimate the number of fine
regular star triangulations (FRST) of three-dimensional reflexive
polytopes \cite{Kreuzer:1998vb}. Each such FRST determines a
smooth weak-Fano toric variety. These varieties
and their number are interesting for at least three reasons: their
anti-canonical hypersurfaces give rise to K3 surfaces that can be used
for six-dimensional string compactifications, they give rise to smooth
F-theory compactifications without non-Higgsable clusters
\cite{Halverson:2016tve}, and they serve as a starting point for
topological transitions from which the ensemble of $10^{755}$ F-theory
geometries arises.

\subsection{Learning Strategy}
Let $\Delta^\circ$ be a $3d$ reflexive polytope. Such an object is the
convex hull of a set of points $\{v\} \subset\bZ^3$ that satisfies the
reflexivity condition, i.e. that the dual polytope
\begin{equation}
  \Delta := \{ m\in \bZ^3 \,\, | \,\, m\cdot v \geq -1 \,\, \forall v \in \Delta^\circ\}
\end{equation}
is itself a lattice polytope. There are $4319$ such
polytopes, classified by Kreuzer and Skarke~\cite{Kreuzer:1998vb}.
A triangulation of $\Delta^\circ$ is
said to be fine, regular, and star if all integral points of $\dc$
are used, the simplicial cones are projections of cones from
an embedding space, and all simplices have the origin as a vertex. We
refer to these as FRSTs.

A weak-Fano toric variety may be associated to each such FRST of a $3d$ reflexive polytope, where $h^{11}(B)= |\{v\}|-3$ is the dimension of
the Dolbeault cohomology group $H^{11}(B)$.\footnote{We remind the reader that we employ the notation $|\{X\}|$ to indicate the cardinality of the set $X$.} This integer measures the number
of independent cohomologically non-trivial $(1,1)$-forms on $B$, or
alternatively (via duality) the number of divisor classes. Such
topological quantities are central to many aspects of the physics of
an F-theory compactification on $B$.

The number of FRSTs
of these polytopes was computed for low $h^{11}(B)$ in~\cite{Halverson:2016tve}, where $B$
is the toric variety associated to the FRST, and estimates were provided for the
remainder based on techniques in the triangulation literature.
Here we instead wish to estimate the number of FRSTs of the $3d$
reflexive polytopes using machine learning.

Our method is to estimate the number of FRSTs of the $3d$ reflexive
polytopes as the product of the number of fine regulation
triangulations (FRTs) of its codimension one faces, which are known as
facets. This was demonstrated to be a good approximation in~\cite{Halverson:2016tve}, and the number of FRTs of the facets was
explicitly computed for $h^{11}(B)\leq 22$, where all $B$ arising from
FRSTs of $3d$ reflexive polytopes have $h^{11}(B)\leq 35$.
In this work, we will utilize machine learning to train a model to predict
the number of FRTs per facet, and then we will use the trained algorithm
to estimate the number of FRSTs of the $3d$ reflexive polytopes.  We
will see that the results are in good agreement with those of~\cite{Halverson:2016tve}, though derived in a different manner.

The vertices of $\Delta^\circ$ determine $\Delta^\circ$, and therefore
its FRSTs may be computed from this data. However, for higher
$h^{11}(B)$ the number of integral points in $\Delta^\circ$ also
increases, which increases the number of FRSTs and therefore also the
likelihood that the computation does not finish in a reasonable amount
of time.  As this occurs, the number of FRTs of each facet also
typically increases.  The number of FRTs of a facet F
increases with its number of points $n_p$, interior points $n_i$,
boundary points $n_b$, and vertices $n_v$, which are of course
related.  To each facet, which is a $2d$ polyhedron, we therefore
associate a $4$-tuple $(n_p,n_i,n_b,n_v)$.
Using machine learning we will train a model $A$ to
predict the number of FRTs of each facet given the $4$-tuple. This gives a chain of operations
\begin{equation}
F \longrightarrow (n_p,n_i,n_b,n_v) \xrightarrow{A} n_T,
\end{equation}
that predicts the $n_T$, the number of FRTs of $F$.
It is obvious that $n_T$ will depend on the $4$-tuple, but the question is to what extent.

We have attempted to choose the training variables wisely based on
knowledge of the dataset. This is supervised learning.

\subsection{Model Evaluation with $10$-fold Cross-Validation}

We begin by utilizing $10$-fold cross-validation to determine which
machine learning algorithm gives rise to the model with the best predictions for $n_T$. There are
two critical considerations that will enter into our analysis. First,
in order to extrapolate to very high $\hoo$ with reasonable confidence,
we would like to train our models for $\hoo< 19$ so that we can test the
trained model on known results for the number of facet FRTs for polytopes with $\hoo=19,20,21$.
We therefore train on data with $\hoo \leq \hoo_{\text{max}}<19$. Second, since
there are very few triangulations for low $\hoo$ and this may negatively affect the model,
we will train on data with $\hoo\geq \hoo_{\text{min}}$. We take $\hoo_{\text{min}}\in \{1,6,10\}$
and $\hoo_{\text{max}}\in \{14,16,18\}$, and therefore we perform a $10$-fold cross
validation on nine different ranges $\hoomin \leq \hoo \leq \hoomax$. 

\begin{table}[t]
  \centering
  \hspace{-.5cm}
  \scalebox{.9}{
    \begin{tabular}{c|c|c|c|c|c|c|c|c|c|}
     \multicolumn{4}{c}{} & \multicolumn{2}{|c|}{Extrap. $h^{11}=19$} & \multicolumn{2}{|c|}{Extrap. $h^{11}=20$} & \multicolumn{2}{|c|}{Extrap. $h^{11}=21$}\\ \hline
    Alg. & $\hoo_{\text{min}}$ & $\hoo_{\text{max}}$ & Train MAPE& MAPE & STDEV & MAPE & STDEV & MAPE & STDEV \\ \hline
LDA&1&14&14.3&28.5&70.0&37.4&85.2&38.0&81.4\\ \hline
KNNR&1&14&4.1&17.4&45.0&19.5&47.2&24.7&53.7\\ \hline
CART&1&14&5.6&18.0&45.3&17.0&39.0&24.3&56.4\\ \hline
NB&1&14&11.2&26.4&58.3&31.9&68.1&35.3&71.7\\ \hline
LDA&1&16&14.5&23.1&68.7&31.9&87.3&30.8&80.4\\ \hline
KNNR&1&16&4.4&15.0&43.9&20.0&63.6&26.6&76.1\\ \hline
CART&1&16&5.6&14.2&42.4&14.5&43.3&20.9&58.5\\ \hline
NB&1&16&10.9&21.3&56.6&28.2&70.7&29.8&70.3\\ \hline
LDA&1&18&15.1&20.9&67.9&28.8&86.8&28.7&79.9\\ \hline
KNNR&1&18&4.7&12.2&33.0&12.0&46.7&19.9&61.0\\ \hline
CART&1&18&6.2&11.7&41.0&11.1&40.5&18.5&53.9\\ \hline
NB&1&18&11.9&19.3&55.6&25.2&69.9&27.4&69.6\\ \hline
LDA&6&14&14.6&26.8&70.0&35.1&85.4&35.7&81.7\\ \hline
KNNR&6&14&4.4&18.4&39.4&21.9&50.1&27.0&59.5\\ \hline
CART&6&14&5.3&18.2&45.6&17.5&40.5&25.0&58.0\\ \hline
NB&6&14&10.7&25.3&58.1&31.6&67.8&34.1&71.5\\ \hline
LDA&6&16&14.8&22.4&68.6&31.1&87.3&30.0&80.3\\ \hline
KNNR&6&16&4.7&13.8&40.7&16.8&55.2&22.9&64.8\\ \hline
CART&6&16&6.0&13.3&41.6&13.4&41.9&19.4&53.3\\ \hline
NB&6&16&11.1&19.9&56.4&27.9&70.7&29.0&70.3\\ \hline
LDA&6&18&14.0&20.4&67.9&28.2&86.7&28.2&80.3\\ \hline
KNNR&6&18&5.1&10.8&38.9&11.1&48.4&18.3&60.0\\ \hline
CART&6&18&6.1&11.8&40.8&10.5&40.0&17.5&54.4\\ \hline
NB&6&18&12.6&18.3&55.6&24.6&69.8&26.0&70.2\\ \hline
LDA&10&14&12.7&16.7&44.1&17.5&47.5&22.8&56.9\\ \hline
KNNR&10&14&5.4&16.9&44.7&19.7&49.1&25.4&58.8\\ \hline
CART&10&14&6.2&16.3&43.9&16.3&44.5&21.6&54.4\\ \hline
NB&10&14&10.3&22.7&58.2&30.7&71.6&32.0&71.9\\ \hline
LDA&10&16&12.9&14.8&42.9&15.6&46.0&21.7&56.4\\ \hline
KNNR&10&16&5.4&14.4&35.5&17.1&59.9&25.1&77.7\\ \hline
CART&10&16&6.3&14.0&42.1&14.2&44.8&21.0&55.2\\ \hline
NB&10&16&10.5&19.8&56.5&28.3&70.7&29.1&70.5\\ \hline
LDA&10&18&12.9&12.6&41.0&12.1&43.5&18.6&54.7\\ \hline
KNNR&10&18&5.9&11.5&38.7&12.4&46.9&19.5&59.2\\ \hline
CART&10&18&7.2&12.4&41.0&11.1&40.4&17.9&53.6\\ \hline
NB&10&18&11.2&17.9&55.5&25.2&69.8&25.9&69.6\\ \hline
    \end{tabular}}
  \caption{Model discrimination and higher $h^{11}$ testing for the number
    of FRTs of $3d$ reflexive polytopes. Train MAPE is the mean average percent
    error of the algorithm of a given type, across the ten training runs, trained on the exact number of FRTs
    for $\hoomin \leq \hoo \leq \hoomax$. The STDEV is the standard deviation of the ten training runs about the MAPE value. Both MAPE and STDEV are presented for model
    predictions for $\hoo=19,20,21>\hoomax$.}  
 \label{tab:3dpolyopt}
 \end{table}

For each, we test four different algorithms:
\begin{itemize}
\item LDA: Linear Discriminant Analysis
\item KNNR: k-Nearest Neighbors Regression
\item CART: Classification and Regression Trees 
\item NB: Naive Bayes,
\end{itemize}
which are described in Section~\ref{sec:mlforst}.
The scoring metric that we use is the mean absolute percent error,
which is defined to be (for $k$-fold validation)
\begin{equation}
\text{MAPE}:= \frac{100}{k} \times \sum_{i=1}^k \left| \frac{A_i-P_i}{A_i}\right|,
\end{equation}
where $P_i$ and $A_i$ are the predicted and actual values for the output; here $n_T$ of the $i^\text{th}$ facet. Finally, for each algorithm trained on the data with $\hoomin \leq \hoo \leq \hoomax$,
we predict $n_T$ for each polytope with $\hoo=19,20,21$ and present the MAPE.

\begin{table}[t]
  \centering
  \hspace{-.5cm}
  \scalebox{.9}{
    \begin{tabular}{c|c|c|c|c|c|c|c|c|c|}
     \multicolumn{4}{c}{} & \multicolumn{2}{|c|}{Extrap. $h^{11}=19$} & \multicolumn{2}{|c|}{Extrap. $h^{11}=20$} & \multicolumn{2}{|c|}{Extrap. $h^{11}=21$}\\ \hline
      Alg. & $\hoo_{\text{min}}$ & $\hoo_{\text{max}}$ & Train MAPE& MAPE & STDEV & MAPE & STDEV & MAPE & STDEV \\ \hline
CART&1&18&6.2&11.9&41.2&11.1&40.5&18.5&53.9\\ \hline
CART&2&18&5.6&11.8&40.5&10.4&39.3&17.7&53.7\\ \hline
CART&3&18&5.7&11.5&40.4&10.3&39.3&17.7&54.5\\ \hline
CART&4&18&5.5&11.2&40.1&10.5&40.0&17.4&54.4\\ \hline
CART&5&18&6.0&12.1&41.0&12.2&41.4&19.3&55.2\\ \hline
CART&6&18&6.2&11.6&40.6&10.5&40.0&17.5&54.4\\ \hline
CART&7&18&5.9&11.5&40.5&10.5&40.0&17.5&54.4\\ \hline
CART&8&18&6.5&11.6&40.5&10.5&40.0&17.6&54.4\\ \hline
CART&9&18&6.8&12.5&41.1&11.6&40.8&18.9&54.2\\ \hline
CART&10&18&7.2&12.1&40.7&11.1&40.4&17.9&53.6 \\ \hline
 \end{tabular}}
\caption{Refinement of CART algorithms for final model selection.}
\label{tab:3dpolyopt2}
\end{table}

The results of this analysis are presented in Table~\ref{tab:3dpolyopt}. The minimal MAPE
for the training set occurs for KNNR with $(\hoomin,\hoomax)=(1,14)$; however, we see that
this case has higher MAPE for $n_T$ of facets at higher $h^{11}$. The lowest MAPE at $\hoo=21$
occurs in CART examples with $(\hoomin,\hoomax)=(6,18)$ and $(\hoomin,\hoomax)=(10,18)$,
with slightly better performance in the former case. In Table \ref{tab:3dpolyopt2} we present
the results of a similar analysis that fixes $\hoomax$ and CART while scanning over $\hoomin$.
The results are similar in all cases, though the MAPE is slightly lower in the case $\hoomin=4$,
which seems to be a point at which the anomalous polytopes at very low $\hoo$ are safely
discarded.

As a result of this analysis, we choose to model the FRTs of facets F of $3d$ reflexive polytopes
via a classification and regression tree (CART) with $\hoomin=4$. We will train the model on
$\hoomin=4\leq \hoo \leq 21$ since the exact number of FRTs at $19\leq h^{11} \leq 21$ should
increase accuracy in the extrapolation to $\hoo > 21$.

\begin{figure}[th]
  \hspace{-1.5cm}
  \includegraphics[scale=.4]{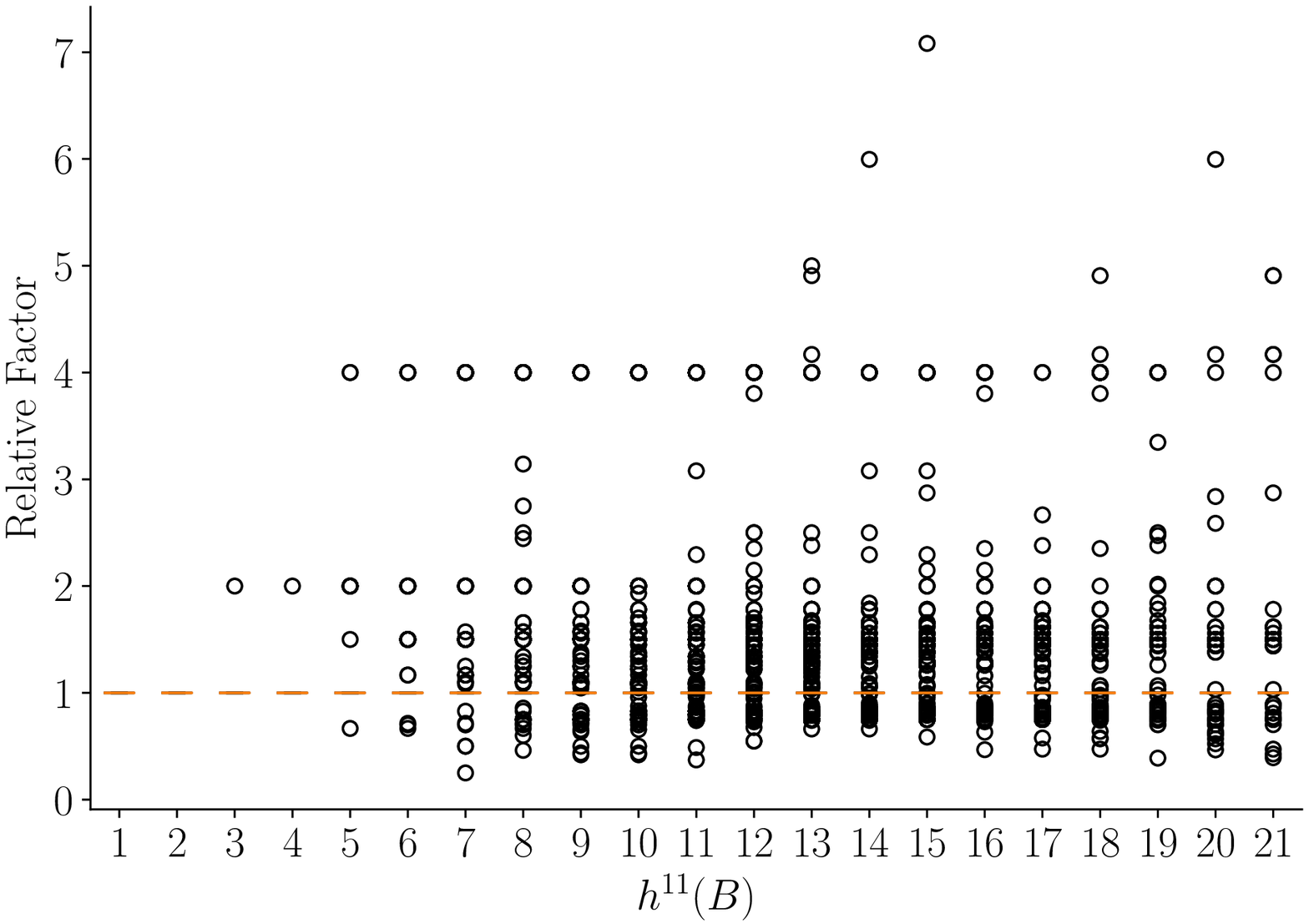}
  \includegraphics[scale=.4]{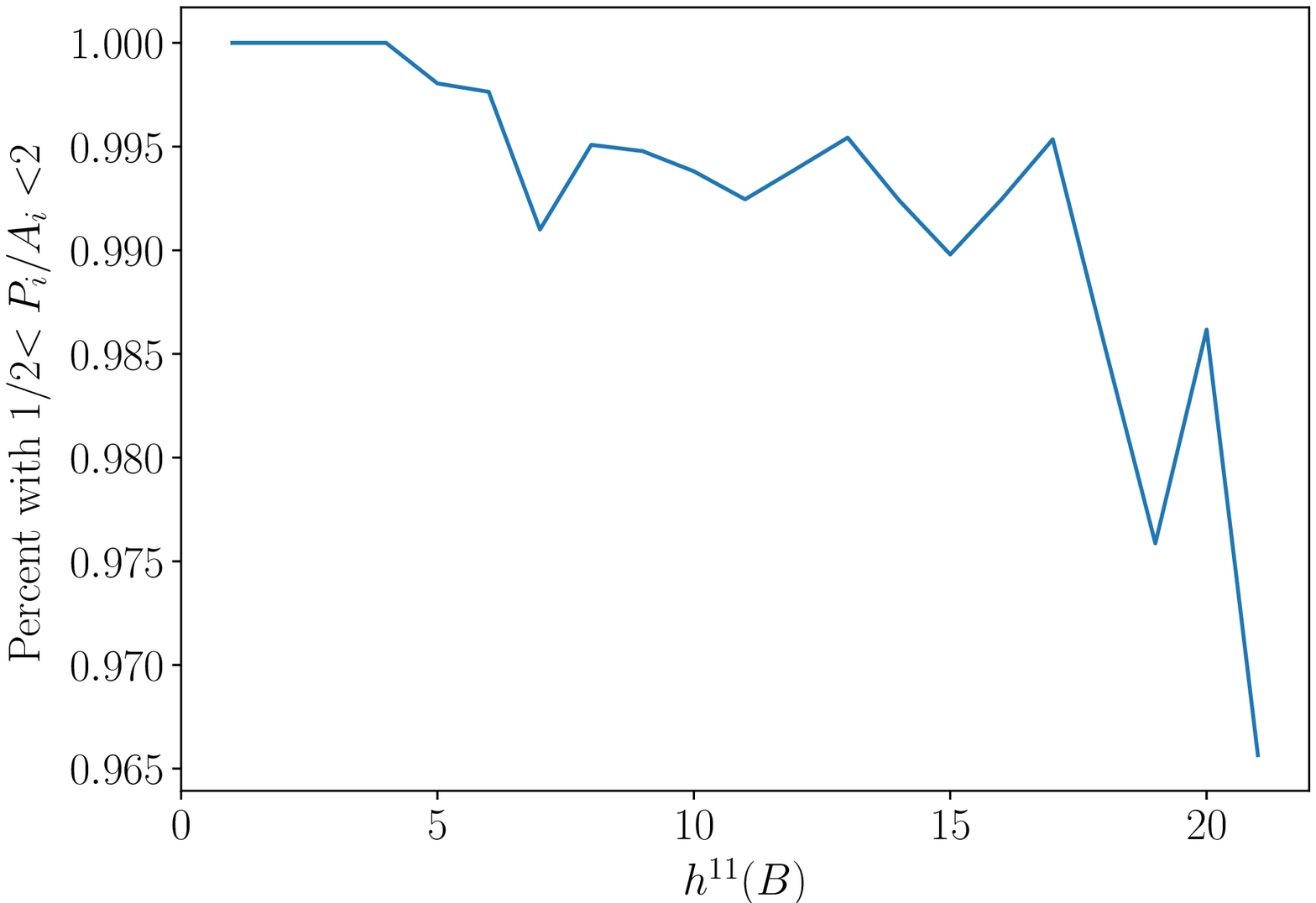}
  \caption{Left: box plot for the relative factor $R_i$, which is the predicted number
    of FRTs of each facet over the actual number. The median, first, and third quartiles are precisely at the desired value $1$,
    though outliers do exist. Right: the percent of facets for which the predicted number
    of FRTs is within a factor of two of the action value.}
  \label{fig:3dpolywhiskerandfactor}
\end{figure}

Training with these parameters using $10$-fold cross validation, we
find that the MAPE is $6.38\pm 1.04\%$; this is on par with the
results of Table \ref{tab:3dpolyopt}. For a broader view of the model
predictions, see Figure \ref{fig:3dpolywhiskerandfactor}. Both
of the plots are a measure of the relative factors
\begin{equation}
R_i := \frac{P_i}{A_i}
\end{equation}
that determine the factor by which the predicted value $P_i$ for $n_T$ of
a facet is off from the actual value $A_i$. We have computed $A_i$ for
all facets in all polytopes up through $\hoo = 21$. The plot on the left
is a box plot of the values of $n_T$ that occur at each respective value
of $h^{11}$. Though a few outlier predictions are off from the actual values
by a factor of $5$ to $7$, note that the orange band denotes the median
which is $1.0$, and the would-be boxes of the box plot are absent since
their boundaries would denote the first and third quartile, both of
which are also $1.0$. The plot on the right computes the percent
with $\frac12 < R_i < 2$; over $96\%$ of the $n_T$ of facets, for all $\hoo$,
are within a factor of $2$ of their actual value.

\begin{figure}[th]
  \centering
  \includegraphics[scale=.5]{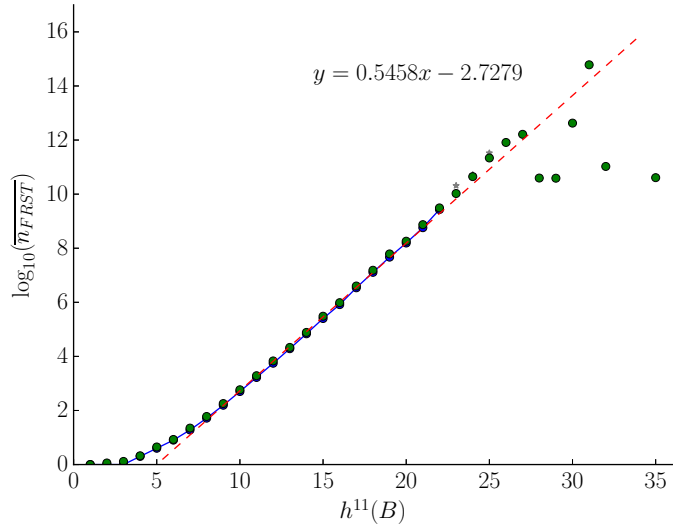}
  \caption{The logarithm of the average number of FRST per polytope. The green dots are
    predictions of our learned model, and the rest of the data is from \cite{Halverson:2016tve}.
    Note the accuracy of model in recovering known results represented by the blue dots
    and grey stars. The erratic behavior for $\hoo \geq 27$ correlates with being
    the tail of the polytope distribution.}
  \label{fig:num3dfrst}
\end{figure}

\begin{figure}[th]
  \hspace{-1.5cm}
  \includegraphics[scale=.4]{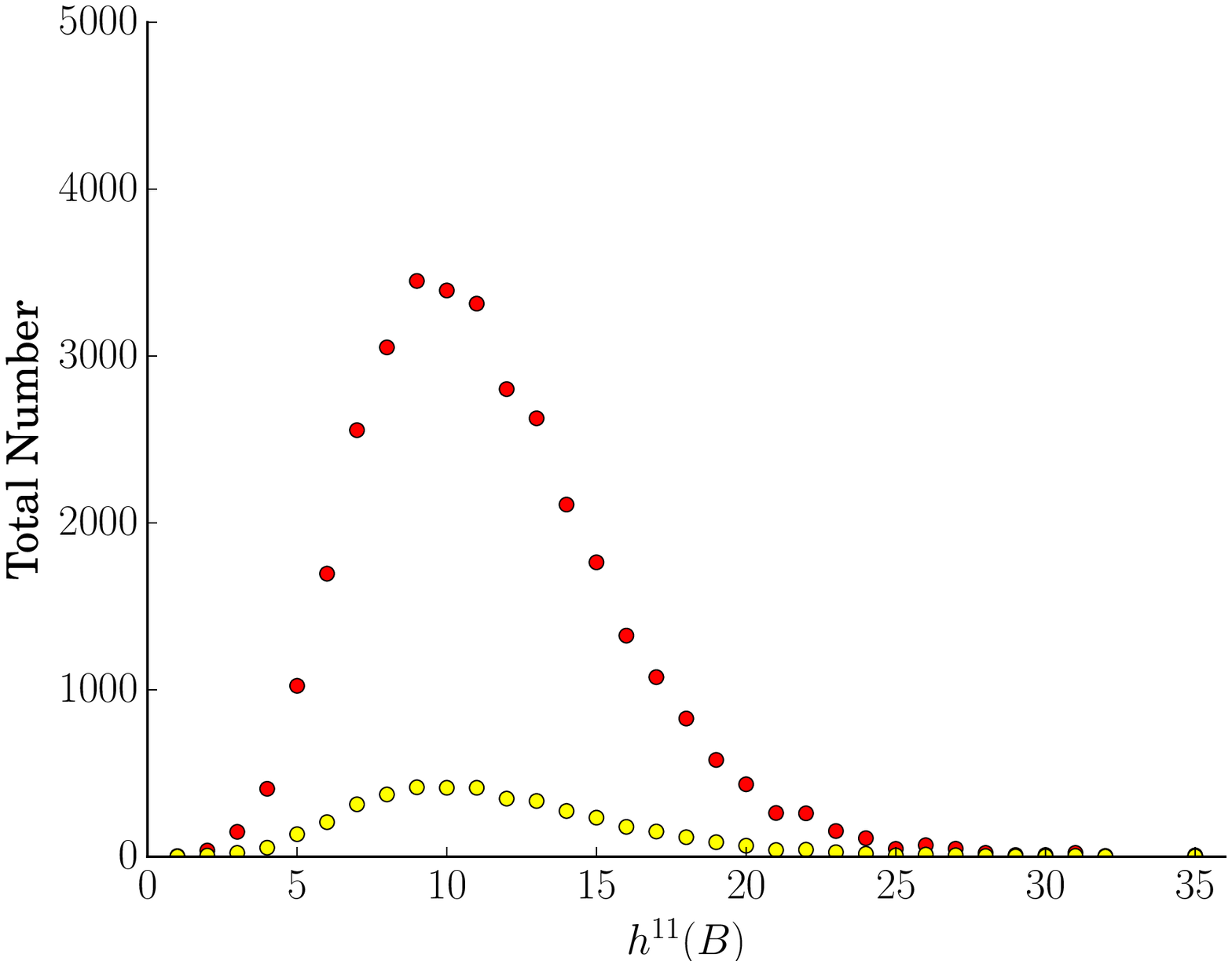}
  \includegraphics[scale=.4]{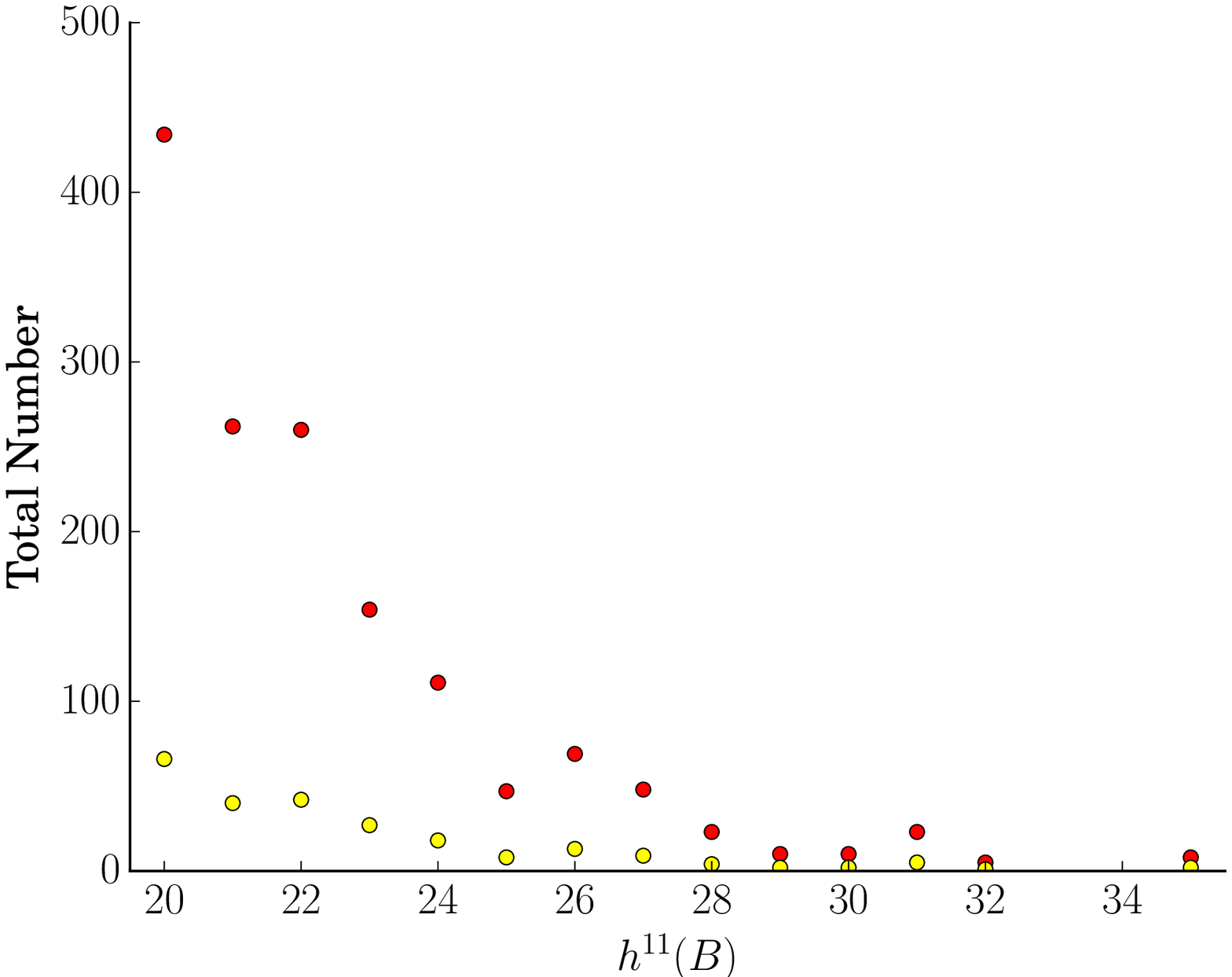}
  \caption{Red (blue) dots are the total number of facets (reflexive polytopes)
    at a given value of $\hoo(B)$. Left: the full plot. Right: magnification of
    the $\hoo \geq 20$ region.}
  \label{fig:num3dfacetsandpolys}
\end{figure}
\subsection{Machine Learning $3d$ Polytope Triangulations}

Given the accuracy of this model, we would now like to compute the
average number of FRSTs per polytope as a function of
$h^{11}(B)$. This was done in \cite{Halverson:2016tve}, and it will be instructive to see
whether machine learning recovers those results, and also where it has
difficulties. Specifically, letting $n_T(F)$ be the number of FRTs of
$F$ computed by our model, we use the approximation discussed in \cite{Halverson:2016tve}
\begin{equation}
n_{FRST}(\Delta^\circ) \simeq \prod_{F\in \Delta^\circ} n_T(F).
\end{equation}
Summing over all polytopes at a given $\hoo$ and averaging, we define
\begin{equation}
  \overline{n_{FRST}}(\hoo) := \frac{\sum_{\Delta^\circ \,\,\text{at}\,\, \hoo} \,\,\,n_{FRST}(\Delta^\circ)}{\sum_{\Delta^\circ \,\,\text{at}\,\, \hoo}\,\,\, 1}.
\end{equation}
This average number of triangulations per polytope at a fixed value of
$\hoo$ is presented in Figure \ref{fig:num3dfrst}. The red line, gray
stars, and blue dots were obtained in \cite{Halverson:2016tve}, where
the gray stars and blue dots were predicted with different
methods. The green dots are the new predictions of our model. The
predictions are so accurate that the green dots are mostly covering
the blue dots, when they exist.

It is also important that our model makes accurate predictions beyond
the data on which it trained. Specifically, it trained on data
with $h^{11}\leq 21$ and the predicted values at $\hoo=22,23,24,25$
are in good agreement with the results of \cite{Halverson:2016tve}. The model also makes
good predictions at $\hoo=26,27$ when comparing to the extrapolation
of the best fit line. We see that our model made six accurate predictions
for $\overline{n_{FRST}}(\hoo)$ for $\hoo$ values that it
was not trained upon. This demonstrates the power of machine learning
to find hidden dependencies in the data that allow for extrapolation
beyond the training set.

Note, however, that the machine learning model
predicts erratic behavior for $\hoo > 27$. While a priori this may
be considered a prediction, the data points are inconsistent with bounds
derived in \cite{Halverson:2016tve} utilizing results in the triangulation literature. Figure
\ref{fig:num3dfacetsandpolys} demonstrates that around this $\hoo$ value
the number of facets and polytopes at a fixed value of $\hoo$ has dropped
significantly relative to lower $\hoo$. This, together with the violation of the analytic bound,
leads to the conclusion that the erratic behavior for
$\hoo > 27$ may be due to being in the tail of the distribution;
see values of $\hoo > 27$ in Figure \ref{fig:num3dfacetsandpolys}.

However, the machine learning model predictions for the bulk of the
polytope distribution picks out the correct line for
$log_{10}(\overline{n_{FRST}})$. This line, when extrapolated to the
highest polytope $\hoo$, which is $\hoo=35$, is in agreement with
the bound
\begin{equation}
5.780\times 10^{14} \lesssim n_{FRST} \lesssim 1.831\times 10^{17},
\end{equation}
which was obtained in \cite{Halverson:2016tve}.

In conclusion, our model that used a $10$-fold cross-validated and
parameter optimized classification and regression tree accurately
predicts the average number of FRSTs in the bulk of the distribution
for the number of polytopes and facets as a function of $\hoo$. It would be interesting
in the future to determine methods that minimize the error in the tail
of the distribution, but we emphasize that the
extrapolated best fit line of the machine learning model prediction
does give $n_{FRST}\sim O(10^{15})-O(10^{16})$ at $h^{11}=35$ as
expected from the analytic results of \cite{Halverson:2016tve}.

\section{Conjecture Generation: Gauge Group Rank in F-theory Ensembles\label{sec:rk}}

In this section, and Section~\ref{sec:e6}, we will generate conjectures related to the
physics of an ensemble of $\frac43 \times 2.96 \times 10^{755}$ F-theory compactifications
that recently appeared in the literature \cite{Halverson:2017ffz}. This ensemble
exhibited algorithmic universality, which is universality derived from a precise
construction algorithm rather than a constructed ensemble. Universal features
included non-Higgsable clusters with extremely large and calculable gauge groups.

\subsection{F-theory Review}

F-theory \cite{Vafa:1996xn,Morrison:1996pp} is a non-perturbative
formulation of the type IIb superstring in which the axiodilaton
$\tau = C_0 + ie^{-\phi}$ varies holomorphically over extra dimensions
of space $B$. This variation is conveniently encoded in the geometry
of a Calabi-Yau elliptic fibration, in which $\tau$ is the complex structure
of the elliptic curve that varies over the base $B$. By a theorem of Nakayama \cite{Nakayama},
every elliptic fibration is birationally equivalent to a Weierstrass model
\begin{equation}
y^2 = x^3 + f x + g,
\end{equation}
where $x,y$ are coordinates on the elliptic curve and $f\in \Gamma(\cO(-4K))$ and
$g \in \Gamma(\cO(-6K))$ are global sections of the states line bundles, with $-K$ the anticanonical class
on $B$. Practically, this simply means that $f$ and $g$ are homogeneous polynomials
in appropriate coordinates on $B$. Seven-branes are localized on the discriminant
locus
\begin{equation}
\Delta = 4f^3 + 27g^2 = 0,
\end{equation}
where the elliptic fiber becomes singular. If $x_i=0$ is a component
of the discriminant locus, then the multiplicity of vanishing
$mult_{x_i}(f,g,\Delta)=(l_i, m_i, n_i)$ of $f,g$ and $\Delta$ along
$x_i=0$ determines (up to monodromy, see e.g. \cite{Anderson:2014gla})
the gauge group on $x_i=0$ according to the Kodaira classification,
which is presented in Table \ref{tab:gauge}. These groups can be understood
via smoothings of the geometry and associated Higgs mechanisms;
for smoothings via K\" ahler resolution and complex structure
deformation see e.g. \cite{Marsano:2011hv,Lawrie:2012gg,Hayashi:2014kca,Braun:2014kla,Braun:2015hkv} and \cite{Grassi:2013kha,Grassi:2014sda,Grassi:2014ffa,Grassi:2016bhs}, respectively.

Typically, the topological structure of the extra spatial dimensions
$B$ forces the existence of non-trivial seven-branes on fixed
divisors. Such seven-branes are referred to as non-Higgsable
seven-branes (NH7), as their inability to move obstructs the Higgs
mechanism that would arise from brane splitting. NH7 usually come in sets,
and they may intersect, forming a non-Higgsable cluster \cite{Morrison:2012np} (NHC).
Mathematically the origin of this mechanism is simple: if the polynomials $f$ and $g$
are chosen to have all possible monomial coefficients non-zero and generic,
then there may be components (factors)
\begin{equation}
f = F \prod_i x_i^{l_i} \qquad g = G \prod x_i^{m_i}, 
\end{equation}
where $F$ and $G$ may themselves be non-trivial polynomials.
If $l_i, m_i >0$, then there is a seven-brane on $x_i=0$ of a type that can
be determined from Table \ref{tab:gauge} and the fact that $n_i = min(3l_i,2m_i)$.
Non-Higgsable clusters are easily exemplified; for example, in a $6d$ F-theory
compactification on the Hirzebruch surface $\mathbb{F}_3$, there is a non-Higgsable
$SU(3)$ on the $-3$ curve. That theory is non-Higgsable because the $SU(3)$ theory
has no matter.

There are a number of interesting recent results about non-Higgsable
clusters and their importance in the $4d$ F-theory landscape. They
exist for generic vacuum expectation values of the complex structure
moduli, and therefore obtaining gauge symmetry does not require moduli
stabilization to fix vacua on subloci in moduli
space~\cite{Grassi:2014zxa}, which can occur at high codimension
\cite{Braun:2014xka,Watari:2015ysa}, though the problem is not as
severe \cite{Halverson:2016tve} as initially thought. Strong coupling
is generic \cite{Halverson:2016vwx}, and some of the structures of the
standard model may arise naturally \cite{Grassi:2014zxa}. They also
arise in the particular geometry $B_{max}$ with the largest known
number of flux vacua \cite{Taylor:2015xtz}, and universally in large
ensembles that have been studied recently
\cite{Halverson:2015jua,Taylor:2015ppa,Halverson:2017ffz}. $4d$ NHC
give rise to interesting features \cite{Morrison:2014lca} (such as
loops and branches) not present in $6d$.  $6d$ NHC have also been
studied extensively
\cite{Morrison:2012np,Morrison:2012js,Taylor:2012dr,Morrison:2014era,Martini:2014iza,Johnson:2014xpa,Taylor:2015isa},
in the context of both $6d$ string universality and $(1,0)$ SCFTs.

\vspace{1cm}
Non-Higgsable clusters can be studied for general bases $B$, see
e.g. \cite{Morrison:2014lca}, but since our examples will be in the case
that $B$ is a toric variety, we will specialize to that case immediately. A
compact toric variety $B$ is specified by a complete fan that is made up
of rays $v_{i}$ that generate cones whose union determines the fan. For
most of what we do in this paper, the $v_i$ will play a starring role and
the cone structure will be less important. For example, for $dim_{\mathbb{C}}(B)=3$ as in our case, the monomials that
may appear in $f$ and $g$ are determined by integral points of the polyhedra
\begin{equation}
\Delta_f = \{m\in \bZ^3 \, | \, m\cdot v_i + 4 \geq 0 \,\, \forall i \} \qquad \qquad
\Delta_g = \{m\in \bZ^3 \, | \, m\cdot v_i + 6 \geq 0 \,\, \forall i \}
\label{eqn:dfdg}
\end{equation}
by the correspondence
\begin{equation}
m_f \in \Delta_f \,\, \mapsto \,\,  \prod_i x_i^{m_f \cdot v_i+4} \qquad 
m_g \in \Delta_g \,\, \mapsto \,\,  \prod_i x_i^{m_g \cdot v_i+6},
\label{eqn:mon}
\end{equation}
which may appear in $f$ and $g$, respectively, and where each $x_i$ is a homogeneous
coordinate on $B$ that is in one to one correspondence with $v_i$. The most general
$f$ and $g$ are therefore of the form
\begin{equation}
f = \sum_{m_f \in \Delta_f} a_f\, \prod_i x_i^{m_f \cdot v_i+4} \qquad g = \sum_{m_g \in \Delta_g} a_g\, \prod_i x_i^{m_g \cdot v_i+6},
\end{equation}
where no restrictions are made on the monomial coefficients. By studying $\Delta_f$ and
$\Delta_g$, it is a straightforward combinatoric exercise to determine the components
(overall factors) of $f$ and $g$, and therefore the non-Higgsable clusters. The mentioned
example of $\mathbb{F}_3$ with a non-Higgsable $SU(3)$, for example, has $\{v_i\}=\{(1,0),(0,1),(-1,0),(-1,-3)\}$. By computing $\Delta_f$ and $\Delta_g$ to construct
the most general $f$ and $g$, one will find a factorization that give the NH7.

\begin{table}[t]
  \centering
\scalebox{1}{\begin{tabular}{|c|c|c|c|c|c|}
\hline
$F_i$ & $l_i$ & $m_i$ & $n_i$ & Sing. & $G_i$ \\ \hline \hline
$I_0$&$\geq $ 0 & $\geq $ 0 & 0 & none & none \\
$I_n$ &0 & 0 & $n \geq 2$ & $A_{n-1}$ & $\gsu(n)$  or $\gsp(\lfloor
n/2\rfloor)$\\
$II$ & $\geq 1$ & 1 & 2 & none & none \\
$III$ &1 & $\geq 2$ &3 & $A_1$ & $\gsu(2)$ \\
$IV$ & $\geq 2$ & 2 & 4 & $A_2$ & $\gsu(3)$  or $\gsu(2)$\\
$I_0^*$&
$\geq 2$ & $\geq 3$ & $6$ &$D_{4}$ & $\gso(8)$ or $\gso(7)$ or $\gg_2$ \\
$I_n^*$&
2 & 3 & $n \geq 7$ & $D_{n -2}$ & $\gso(2n-4)$  or $\gso(2n -5)$ \\
$IV^*$& $\geq 3$ & 4 & 8 & $E_6$ & $\ge_6$  or $\gf_4$\\
$III^*$&3 & $\geq 5$ & 9 & $E_7$ & $\ge_7$ \\
$II^*$& $\geq 4$ & 5 & 10 & $E_8$ & $\ge_8$ \\ \hline
\end{tabular}}
\caption{Kodaira fiber $F_i$, singularity, and gauge group $G_i$ on
the seven-brane at $x_i=0$.}
\label{tab:gauge}
\end{table}

\subsection{A Large Ensemble of F-theory Geometries}
Since it is central to our machine learning analysis, let us review
the construction of~\cite{Halverson:2017ffz} that gives rise to
$\frac43 \times 2.96 \times 10^{755}$ F-theory geometries.

The construction performs a large number of topological transitions
away from an initial algebraic threefold $B_i$, generating a large
number of other threefolds, and then uses them as F-theory bases
$B$. The Calabi-Yau elliptic fibrations over these $B$ form one
connected moduli space with many branches, and all of the $B$ are
different algebraic varieties.

Specifically, the initial threefold $B_i$ is a  smooth weak-Fano toric
threefold, which can be specified by a fine regular star
triangulation (FRST) of a three-dimensional reflexive polytope $\dc$; using
machine learning to estimate the number of such triangulations was the subject
of Section~\ref{sec:3dPoly}. The fan associated to $B_i$ is composed of 2-cones
and 3-cones that appear as edges and faces (triangles) on the real
codimension one faces of $\dc$ in $\bZ^3$, which are known as facets.
The number of edges and faces in the facet are determined by the number
of boundary and interior points in the facet, and these numbers are triangulation
independent. The largest facets appear in the ensemble of $4319$ reflexive polytopes
are 
\begin{center}
  \begin{tikzpicture}[scale=.8]
\draw[thick,color=Black] (0,0) -- (3,0) -- (0,3) -- cycle;
\draw[thick,color=Black] (0,.5) -- (2.5,.5);
\draw[thick,color=Black] (0,1) -- (2,1);
\draw[thick,color=Black] (0,1.5) -- (1.5,1.5);
\draw[thick,color=Black] (0,2) -- (1,2);
\draw[thick,color=Black] (0,2.5) -- (.5,2.5);
\draw[thick,color=Black] (.5,0) -- (.5,2.5);
\draw[thick,color=Black] (1,0) -- (1,2);
\draw[thick,color=Black] (1.5,0) -- (1.5,1.5);
\draw[thick,color=Black] (2,0) -- (2,1);
\draw[thick,color=Black] (2.5,0) -- (2.5,.5);
\draw[thick,color=Black] (0,2) -- (.5,2.5);
\draw[thick,color=Black] (0,1) -- (1,2);
\draw[thick,color=Black] (0,0) -- (1.5,1.5);
\draw[thick,color=Black] (1,0) -- (2,1);
\draw[thick,color=Black] (2,0) -- (2.5,.5);
\draw[thick,color=Black] (0,1.5) -- (.5,2);
\draw[thick,color=Black] (0,.5) -- (1,1.5);
\draw[thick,color=Black] (.5,0) -- (1.5,1);
\draw[thick,color=Black] (1.5,0) -- (2,.5);
\fill (0,0) circle (.5mm); \fill (0,.5) circle (.5mm); \fill (0,1) circle (.5mm);
\fill (0,1.5) circle (.5mm); \fill (0,2) circle (.5mm); \fill (0,2.5) circle (.5mm);
\fill (0,3) circle (.5mm);
\fill (.5,0) circle (.5mm); \fill (.5,.5) circle (.5mm); \fill (.5,1) circle (.5mm);
\fill (.5,1.5) circle (.5mm); \fill (.5,2) circle (.5mm); \fill (.5,2.5) circle (.5mm);
\fill (1,0) circle (.5mm); \fill (1,.5) circle (.5mm); \fill (1,1) circle (.5mm);
\fill (1,1.5) circle (.5mm); \fill (1,2) circle (.5mm); 
\fill (1.5,0) circle (.5mm); \fill (1.5,.5) circle (.5mm); \fill (1.5,1) circle (.5mm);
\fill (1.5,1.5) circle (.5mm); 
\fill (2,0) circle (.5mm); \fill (2,.5) circle (.5mm); \fill (2,1) circle (.5mm);
\fill (2.5,0) circle (.5mm); \fill (2.5,.5) circle (.5mm);
\fill (3,0) circle (.5mm);
\draw[thick,color=Black] (1,3) -- (7,3) -- (7,0) -- cycle;
\draw[thick,color=Black] (3,2) -- (7,2);
\draw[thick,color=Black] (5,1) -- (7,1);
\draw[thick,color=Black] (6.5,1) -- (6.5,3);
\draw[thick,color=Black] (6,1) -- (6,3);
\draw[thick,color=Black] (5.5,1) -- (5.5,3);
\draw[thick,color=Black] (5,1) -- (5,3);
\draw[thick,color=Black] (4.5,2) -- (4.5,3);
\draw[thick,color=Black] (4,2) -- (4,3);
\draw[thick,color=Black] (3.5,2) -- (3.5,3);
\draw[thick,color=Black] (3,2) -- (3,3);
\draw[thick,color=Black] (7,1) -- (6,3);
\draw[thick,color=Black] (6.5,1) -- (5.5,3);
\draw[thick,color=Black] (6,1) -- (5,3);
\draw[thick,color=Black] (5.5,1) -- (4.5,3);
\draw[thick,color=Black] (5,1) -- (4,3);
\draw[thick,color=Black] (7,2) -- (6.5,3);
\draw[thick,color=Black] (4,2) -- (3.5,3);
\draw[thick,color=Black] (3.5,2) -- (3,3);
\draw[thick,color=Black] (7,0) -- (6.5,1);
\draw[thick,color=Black] (7,0) -- (6,1);
\draw[thick,color=Black] (7,0) -- (5.5,1);
\draw[thick,color=Black] (5,1) -- (4.5,2);
\draw[thick,color=Black] (5,1) -- (4,2);
\draw[thick,color=Black] (5,1) -- (3.5,2);
\draw[thick,color=Black] (3,2) -- (2.5,3);
\draw[thick,color=Black] (3,2) -- (2,3);
\draw[thick,color=Black] (3,2) -- (1.5,3);
\fill (1,3) circle (.5mm); \fill (1.5,3) circle (.5mm); \fill (2,3) circle (.5mm); \fill (2.5,3) circle (.5mm);
\fill (3,3) circle (.5mm); \fill (3.5,3) circle (.5mm); \fill (4,3) circle (.5mm); \fill (4.5,3) circle (.5mm);
\fill (5,3) circle (.5mm); \fill (5.5,3) circle (.5mm); \fill (6,3) circle (.5mm); \fill (6.5,3) circle (.5mm);
\fill (7,3) circle (.5mm);
\fill (7,2) circle (.5mm);
\fill (7,1) circle (.5mm);
\fill (7,0) circle (.5mm);
\fill (6.5,1) circle (.5mm);
\fill (6,1) circle (.5mm);
\fill (5.5,1) circle (.5mm);
\fill (5,1) circle (.5mm);
\fill (6.5,2) circle (.5mm);
\fill (6,2) circle (.5mm);
\fill (5.5,2) circle (.5mm);
\fill (5,2) circle (.5mm);
\fill (4.5,2) circle (.5mm);
\fill (4,2) circle (.5mm);
\fill (3.5,2) circle (.5mm);
\fill (3,2) circle (.5mm);,
\end{tikzpicture}
\end{center}
and this is the sort of picture the reader should have in mind when we refer to
building structure on the ``ground''; this is the ground.

The topological transitions are smooth blow-ups along curves or
points. In the former case, two $3$-cones labeled by their generators
$(v_1,v_2,v_3)$ an $(v_4,v_2,v_3)$ are replaced by $(v_1,v_2,v_e)$,
$(v_1,v_3,v_e)$, $(v_4,v_2,v_e)$, $(v_4,v_3,v_e)$, where $v_e=v_2+v_3$ lies
above the original facet in which $v_1,v_2,v_3,v_4$ lie. The process can
be iterated multiple times, for example doing a similar subdivision that
adds new rays $v_f = v_e + v_2 = v_1 + 2 v_2$ and $v_g = v_e + v_1 = 2 v_1 + v_2$.
The new structure could be visualized in three dimensions as
\begin{center} 
\begin{tikzpicture}[scale=.85]
 \draw[thick,color=Black] (.25,.5) --
(-.25,.5);\draw[thick,dash pattern={on 1pt off 1pt},color=ForestGreen] (-.25,.5)--(-.25,1.5)--(0,1)--(.25,1.5)--(.25,.5); \fill (0,0) circle (.5mm); \fill (.25,.5) circle (.5mm); \fill
(-.25,.5) circle (.5mm); \fill (0,1) circle (.5mm); \fill (-.25,1.5) circle
(.5mm); \fill (.25,1.5) circle (.5mm); \draw (0,0) -- (.25,.5); \draw (0,0) --
(-.25,.5); \draw (0,0) -- (0,1); \draw (0,0) -- (-.25,1.5); \draw (0,0) --
(.25,1.5); \node at (-.55,.5) {$v_2$}; \node at (.55,.5) {$v_3$}; \node at
(0,-.25) {$0$}; \node at (0,1.25) {$2$}; \node at (-.45,1.6) {$3$}; \node at
(.45,1.6) {$3$}; 
\end{tikzpicture} 
\end{center}
where the solid line in between $v_2$ and $v_3$ is the edge on the
facet, or ground, above which the new rays have been added; we may
refer to edges or faces above which new rays are added as \emph{patches}
on the ground. This structure is
the result of a sequence of blowups, and rather than continually
saying ``sequence of blowups'' we will instead refer to the rays
and cone structure associated to the sequence of blow-ups
as \emph{trees}, as suggested naturally by the image, where the dashed
green lines denote new edges above the original patch on the ground. Any new ray $v$ in the fan associated to $B$ may be
written as a linear combination $v=av_1 + bv_2 + c v_3$ if it is above
a face with vertices $v_1,v_2,v_3$ or as $v=av_1+bv_2$ if it is above
an edge with vertices $v_1,v_2$. As the sequence of blow-ups are
trees, we will refer to the new rays $v$ in the tree as its
\emph{leaves}, each of which has a \emph{height} $h=a+b+c$ or $h=a+b$
depending on whether it is above a face or an edge. The numbers in the
picture are the heights of the leaves, and the height measures the
distance of the ground. We will interchangeably refer to $h=1$ leaves
as roots, or leaves on the ground, as with these definitions the $h=1$
leaves were already in the original reflexive polytope $\dc$.  The
height of a tree is defined to be the height of its highest leaf.

Trees built above faces will be referred to as \emph{face trees}, and
those above edges will be referred to as \emph{edge trees}. It is
possible to classify the number of face trees and edge trees with
a fixed maximal height $h\leq N$.
It is convenient to view the edges and trees face on, i.e. with the
leaves projected onto the edge from above. In this case the edge on the ground appears as
\begin{center}
\begin{tikzpicture}
\draw[thick,color=Black] (0,0) -- (1,0);
\fill (0,0) circle (.5mm);
\fill (1,0) circle (.5mm);
\node at (0,.3) {$v_2$};
\node at (1,.3) {$v_3$};
\node at (0,-.3) {$1$};
\node at (1,-.3) {$1$};
\end{tikzpicture}
\end{center}
with the edge vertices and their heights labeled.
Adding $v_1+v_2$ subdivides the
edge, and further subdividing, dropping vertex labels, gives
\begin{center}
\begin{tikzpicture}
\draw[thick,color=Black] (0,0) -- (1,0);
\fill (0,0) circle (.5mm);
\fill (1,0) circle (.5mm);
\node at (0,-.3) {$1$};
\node at (1,-.3) {$1$};
\draw[thick,->] (1.25,0) -- (1.75,0);
\draw[thick,dash pattern={on 1pt off 1pt},color=ForestGreen] (2,0) -- (3,0);
\fill (2,0) circle (.5mm);
\node at (2,-.3) {$1$};
\fill (2.5,0) circle (.5mm);
\node at (2.5,-.3) {$2$};
\fill (3,0) circle (.5mm);
\node at (3,-.3) {$1$};
\draw[thick,->] (3.25,.1) -- (3.75,.38);
\draw[thick,->] (3.25,-.1) -- (3.75,-.38);
\draw[thick,->] (5.25,.38) -- (5.75,.1);
\draw[thick,->] (5.25,-.38) -- (5.75,-.1);
\draw[thick,dash pattern={on 1pt off 1pt},color=ForestGreen] (4,.5) -- (5,.5);
\fill (4,.0+.5) circle (.5mm);
\node at (4,-.3+.5) {$1$};
\fill (4.5,0+.5) circle (.5mm);
\node at (4.5,-.3+.5) {$2$};
\fill (4.75,0+.5) circle (.5mm);
\node at (4.75,-.3+.5) {$3$};
\fill (5,0+.5) circle (.5mm);
\node at (5,-.3+.5) {$1$};
\draw[thick,dash pattern={on 1pt off 1pt},color=ForestGreen] (4,-.5) -- (5,-.5);
\fill (4,.0-.5) circle (.5mm);
\node at (4,-.3-.5) {$1$};
\fill (4.5,0-.5) circle (.5mm);
\node at (4.5,-.3-.5) {$2$};
\fill (4.25,0-.5) circle (.5mm);
\node at (4.25,-.3-.5) {$3$};
\fill (5,0-.5) circle (.5mm);
\node at (5,-.3-.5) {$1$};
\draw[thick,->] (5.25,.38) -- (5.75,.1);
\draw[thick,->] (5.25,-.38) -- (5.75,-.1);
\draw[thick,dash pattern={on 1pt off 1pt},color=ForestGreen] (6,0) -- (7,0);
\fill (6,.0) circle (.5mm);
\node at (6,-.3) {$1$};
\fill (6.25,0) circle (.5mm);
\node at (6.25,-.3) {$3$};
\fill (6.5,0) circle (.5mm);
\node at (6.5,-.3) {$2$};
\fill (6.75,0) circle (.5mm);
\node at (6.75,-.3) {$3$};
\fill (7,0) circle (.5mm);
\node at (7,-.3) {$1$};
\end{tikzpicture}
\end{center}
which are all of the $h\leq 3$ edge trees. Similarly, the face trees may be viewed
face on, and beginning with a face the first blowup gives
\begin{center}
\begin{tikzpicture}[scale=0.9, every node/.style={scale=0.9}]
\draw[thick,color=Black] (90:.75) -- (90+120:.75) -- (90+120+120:.75) -- cycle;
\fill (90:.75) circle (.5mm);
\fill (90+120:.75) circle (.5mm);
\fill (90+240:.75) circle (.5mm);
\node at (90:1) {$1$}; \node at (90+120:1) {$1$}; \node at (90+240:1) {$1$};
\draw[thick,->] (1.25,.1) -- (1.75,.1);
\begin{scope}[xshift=3cm]
\draw[thick,dash pattern={on 1pt off 1pt},color=ForestGreen] (90:.75) -- (0,0);
\draw[thick,dash pattern={on 1pt off 1pt},color=ForestGreen] (90+120:.75) -- (0,0);
\draw[thick,dash pattern={on 1pt off 1pt},color=ForestGreen] (90+240:.75) -- (0,0);
\fill (0,0) circle (.5mm);
\draw[thick,color=Black] (90:.75) -- (90+120:.75) -- (90+120+120:.75) -- cycle;
\fill (90:.75) circle (.5mm);
\fill (90+120:.75) circle (.5mm);
\fill (90+240:.75) circle (.5mm);
\node at (90:1) {$1$}; \node at (90+120:1) {$1$}; \node at (90+240:1) {$1$};
\node at (0,-.2) {$3$};
\end{scope}
\end{tikzpicture}
\end{center}
which are the only two $h\leq 3$ face trees. The green dotted lines in both
pictures denote edges that are above the original ground, due to at least one of the
leaves on the edge having $h>1$, i.e. being above the ground.

Also critical to the construction is a bound $h\leq 6$ on all trees. This bound is sufficient,
but not necessary, to avoid a pathology known as a $(4,6)$ divisor that is not allowed
in a consistent F-theory compactification; see the appendix of~\cite{Halverson:2017ffz}
for an in-depth discussion. Given this bound, it is pertinent to classify all $h\leq 6$
face trees and edge trees. Their number, for all $3\leq N \leq 6$, is
\begin{center}
  \begin{tabular}{|c|c|c|}
\hline
$N$ & \# Edge Trees & \# Face Trees \\ \hline
$3$ & $5$ & $2$\\
$4$ & $10$ & $17$\\
$5$ & $50$ & $4231$ \\
$6$ & $82$ & $41,873,645$\\ \hline
\end{tabular}
\end{center}
and these numbers enter directly into the combinatorics that generate the large ensemble.

The ensemble $\sdc$ associated to a $3d$ reflexive polytope is defined as follows. First,
pick a fine regular star triangulation of $\dc$, denoted $\mathcal{T}(\Delta^\circ)$. Add one of the
$41,873,645$ face trees to each face of $\mathcal{T}(\Delta^\circ)$, and one of the $82$ edge
trees to each edge of $\mathcal{T}(\Delta^\circ)$. 
The size of $S_{\Delta^\circ}$ is 
\begin{equation}
|S_{\Delta^\circ}| = 82^{\# \tilde E \, \text{on} \, \mathcal{T}(\Delta^\circ)} \times (41,873,645)^{\# \tilde F \, \text{on} \, \mathcal{T}(\Delta^\circ)}\,, 
\end{equation}
where $\# \tilde E$ and $\# \tilde F $ are the number of edges and faces on $\mathcal{T}(\Delta^\circ)$, which are triangulation-independent and are entirely determined by $\Delta^\circ$~\cite{DeLoera:2010:TSA:1952022}.

Two of the $3d$ reflexive polytopes give a far larger number $|S_{\Delta^\circ}|$ than
all of the others combined. These
polytopes are the the convex hulls $\Delta_i^\circ := \text{Conv}(S_i), i=1,2$ of the
vertex sets
\begin{align} 
S_1 &= \{ (-1,-1,-1),(-1,-1,5),(-1,5,-1),(1,-1,-1)\}\, , \nonumber \\
  S_2 &= \{ (-1,-1,-1),(-1,-1,11),(-1,2,-1),(1,-1,-1)\}. \nonumber
\end{align}
Surprisingly,
$\mathcal{T}(\doc)$ and $\mathcal{T}(\dtc)$ have the same number of edges and faces.
Their largest facets were the ones previously displayed, and they have
$\# \tilde E = 63$ and $\# \tilde F=36$. This gives
\begin{equation}
|\sdoc| = \frac{2.96}{3} \times 10^{755} \qquad |\sdtc| = 2.96 \times 10^{755},
\label{eqn:sdocsdtccounts}
\end{equation}
where the factor of $1/3$ is due to a particular $\bZ_3$ rotation that gives an
equivalence of toric varieties;
see the appendix of~\cite{Halverson:2017ffz} for a discussion.
All of the other polytopes $\Delta^\circ$ contribute negligibly,
yielding
$|S_{\Delta^\circ}| \leq 3.28\times 10^{692}$ 
configurations. This gives
\begin{equation}
\text{\# 4d F-theory Geometries} \geq \frac43 \times 2.96 \times 10^{755},
\end{equation}
which is a lower bound for a number of reasons discussed in~\cite{Halverson:2017ffz}.

\vspace{1cm}
We end this section with a critical technical point. Previously we described how to read off
the non-Higgsable gauge group for fixed base $B$ by constructing the $\Delta_f$ and $\Delta_g$
polytopes \eqref{eqn:dfdg}, their associated monomials \eqref{eqn:mon}, and from them
the most general possible $f$ and $g$.

We have just introduced a large number of topological transitions
$B\to B'$, and via these transitions the gauge groups on various
leaves may change. The minimal transitions $B\to B'$ arise as from a
single blow-up, which adds an exceptional divisor $x_e=0$ and new ray
$v_e=0$ that wasn't present in the set of rays associated to
$B$. Adding this new ray means that in \eqref{eqn:dfdg} there is an
additional upper half plane condition that must be satisfied. If these upper half planes
slice across $\Delta_f$ and $\Delta_g$, they are changed
\begin{equation}
  \Delta_f \mapsto \Delta_f' \qquad \qquad \Delta_g \mapsto \Delta_g',
  \label{eqn:delchange}
\end{equation}
where $\Delta_f'$ $(\Delta_g'$) contains all points of $\Delta_f$ ($\Delta_g$)
except those removed by the new upper half plane $m\cdot v_e + 4 \geq 0$
($m\cdot v_e + 6\geq 0$).
In such a case we say that the points $m_f \in \Delta_f$
($m_g \in \Delta_g$) that are removed by the process are ``chopped off,'' since the
upper half plane condition forms the new polytope by slicing across the old one and
removing those points.

More generally the process $B\to B'$ may be a sequence of transitions, where the full
sequence adds a tree. In that case there will be as many new upper half planes as
there are new leaves in the tree, and each may chop points out of the original
$\Delta_f$ ($\Delta_g$) to form $\Delta_f'$ ($\Delta_g'$).

The critical physical point is that, in doing transitions $B\to B'$ that chop out
monomials from $\Delta_f$ and $\Delta_g$ to arrive at $\Delta_f'$ and $\Delta_g'$,
one must redo the gauge group analysis, and the chopping procedure may change the
gauge group on vertices present in both $B$ and $B'$. This is absolutely central to the physics
of the construction, as e.g. for the initial $B_i$ the $v\in \dc$ have no gauge group,
but they quickly obtain gauge groups once trees are added. 

\subsection{Data Generation from Random Samples}

\begin{table}
  \centering
  \begin{tabular}{|ccc|ccc|}
    \hline $v_1$ & $v_2$ & $v_3$ & $v_1$ & $v_2$ & $v_3$ \\ \hline
$(-1, -1, -1)$&$(-1, -1, 0)$&$(-1, 0, -1)$&$(-1, -1, -1)$&$(-1, -1, 0)$&$(-1, 0, -1)$\\
$(-1, -1, -1)$&$(-1, -1, 0)$&$(0, -1, -1)$&$(-1, -1, -1)$&$(-1, 0, -1)$&$(0, -1, -1)$\\
$(-1, -1, -1)$&$(-1, 0, -1)$&$(0, -1, -1)$&$(-1, -1, 5)$&$(-1, -1, 4)$&$(0, -1, 2)$\\
$(-1, -1, 5)$&$(-1, -1, 4)$&$(-1, 0, 4)$&$(-1, 5, -1)$&$(-1, -1, 0)$&$(-1, 0, 0)$\\
$(-1, -1, 5)$&$(-1, -1, 4)$&$(0, -1, 2)$&$(-1, 5, -1)$&$(-1, 4, -1)$&$(0, 2, -1)$\\
$(-1, -1, 5)$&$(-1, 0, 4)$&$(0, -1, 2)$&$(-1, 5, -1)$&$(-1, -1, 4)$&$(-1, 4, 0)$\\
$(-1, 5, -1)$&$(-1, -1, 0)$&$(-1, 0, 0)$&$(-1, 5, -1)$&$(-1, 0, 0)$&$(-1, 1, 0)$\\
$(-1, 5, -1)$&$(-1, -1, 0)$&$(-1, 4, -1)$&$(-1, 5, -1)$&$(-1, 0, 1)$&$(-1, 3, 0)$\\
$(-1, 5, -1)$&$(-1, 4, -1)$&$(0, 2, -1)$&$(-1, 5, -1)$&$(-1, 0, 2)$&$(-1, 3, 0)$\\
$(-1, 5, -1)$&$(-1, -1, 4)$&$(-1, 0, 3)$&$(-1, 5, -1)$&$(-1, 1, 0)$&$(-1, 2, 0)$\\
$(-1, 5, -1)$&$(-1, -1, 4)$&$(-1, 4, 0)$&$(1, -1, -1)$&$(-1, -1, 0)$&$(0, -1, -1)$\\
$(-1, 5, -1)$&$(-1, 4, 0)$&$(0, 2, -1)$&$(1, -1, -1)$&$(-1, 0, -1)$&$(0, 0, -1)$\\
$(-1, 5, -1)$&$(-1, 0, 0)$&$(-1, 1, 0)$&$(1, -1, -1)$&$(-1, 2, -1)$&$(0, 0, -1)$\\
$(-1, 5, -1)$&$(-1, 0, 1)$&$(-1, 2, 0)$&$(1, -1, -1)$&$(-1, 4, -1)$&$(0, 1, -1)$\\
$(-1, 5, -1)$&$(-1, 0, 1)$&$(-1, 3, 0)$&$(-1, -1, 2)$&$(-1, 0, 1)$&$(-1, 1, 1)$\\
$(-1, 5, -1)$&$(-1, 0, 2)$&$(-1, 2, 1)$&$(-1, -1, 2)$&$(-1, 0, 2)$&$(-1, 1, 1)$\\
$(-1, 5, -1)$&$(-1, 0, 2)$&$(-1, 3, 0)$&$(1, -1, -1)$&$(-1, -1, 4)$&$(0, -1, 1)$\\
$(-1, 5, -1)$&$(-1, 0, 3)$&$(-1, 1, 2)$&$(1, -1, -1)$&$(-1, 0, 4)$&$(0, -1, 2)$\\
$(-1, 5, -1)$&$(-1, 1, 0)$&$(-1, 2, 0)$&$(1, -1, -1)$&$(-1, 2, 2)$&$(0, 0, 1)$\\
$(-1, 5, -1)$&$(-1, 1, 2)$&$(-1, 2, 1)$&$(1, -1, -1)$&$(-1, 4, 0)$&$(0, 1, 0)$\\
$(1, -1, -1)$&$(-1, -1, 0)$&$(0, -1, -1)$&$(-1, 0, 1)$&$(-1, 1, 1)$&$(-1, 3, 0)$\\
$(1, -1, -1)$&$(-1, 0, -1)$&$(0, -1, -1)$&$(-1, -1, 0)$&$(-1, -1, 1)$&$(-1, 0, 0)$\\
$(1, -1, -1)$&$(-1, 0, -1)$&$(0, 0, -1)$&$(-1, -1, 0)$&$(-1, 0, -1)$&$(-1, 1, -1)$\\
$(1, -1, -1)$&$(-1, -1, 0)$&$(0, -1, 0)$&$(-1, -1, 0)$&$(-1, 1, -1)$&$(-1, 2, -1)$\\
$(1, -1, -1)$&$(-1, 2, -1)$&$(0, 0, -1)$&$(-1, -1, 0)$&$(-1, 2, -1)$&$(-1, 3, -1)$\\
$(1, -1, -1)$&$(-1, 2, -1)$&$(0, 1, -1)$&$(-1, -1, 0)$&$(-1, 3, -1)$&$(-1, 4, -1)$\\
$(1, -1, -1)$&$(-1, 4, -1)$&$(0, 1, -1)$&$(-1, -1, 1)$&$(-1, -1, 2)$&$(-1, 0, 1)$\\
$(1, -1, -1)$&$(-1, 4, -1)$&$(0, 2, -1)$&$(-1, -1, 1)$&$(-1, 0, 0)$&$(-1, 1, 0)$\\
$(-1, -1, 2)$&$(-1, 0, 1)$&$(-1, 1, 1)$&$(-1, -1, 1)$&$(-1, 1, 0)$&$(-1, 2, 0)$\\
$(1, -1, -1)$&$(-1, -1, 2)$&$(0, -1, 0)$&$(-1, -1, 2)$&$(-1, -1, 3)$&$(0, -1, 1)$\\
$(-1, -1, 2)$&$(-1, 0, 2)$&$(-1, 1, 1)$&$(-1, -1, 3)$&$(-1, -1, 4)$&$(0, -1, 1)$\\
$(1, -1, -1)$&$(-1, -1, 2)$&$(0, -1, 1)$&$(-1, -1, 3)$&$(-1, 0, 3)$&$(-1, 1, 2)$\\
$(1, -1, -1)$&$(-1, -1, 4)$&$(0, -1, 1)$&$(-1, -1, 4)$&$(-1, 0, 4)$&$(-1, 1, 3)$\\
$(1, -1, -1)$&$(-1, -1, 4)$&$(0, -1, 2)$&$(-1, -1, 4)$&$(-1, 1, 3)$&$(-1, 2, 2)$\\
$(1, -1, -1)$&$(-1, 0, 4)$&$(0, -1, 2)$&$(-1, -1, 4)$&$(-1, 2, 2)$&$(-1, 3, 1)$\\
$(1, -1, -1)$&$(-1, 0, 4)$&$(0, 0, 1)$&$(-1, -1, 4)$&$(-1, 3, 1)$&$(-1, 4, 0)$ \\ \hline
  \end{tabular}
  \caption{The three-dimensional cones of the pushing triangulation $\cT_p$ of
    $\doc$.}
  \label{tab:pushing}
\end{table}

In this section and Section~\ref{sec:e6} we will utilize random
samples to generate data that is studied via machine learning.
The random samples are generated as follows.

All samples in this
paper focus on the ensemble $\sdoc$, where the $B\in \sdoc$ are forests
of trees that are built on the $3d$ reflexive polytope $\sdoc$. In future
work, it would be interesting to study similar issues in the other
large ensemble, $\sdtc$.

Furthermore, all samples in this paper are built on top of a particular
triangulation $\cT_p$ of $\doc$, the so-called pushing triangulation,
which exists for any reflexive polytope.
For a definition of the pushing triangulation, see~\cite{DeLoera:2010:TSA:1952022}. For our purposes it suffices to simply list the three-dimensional cones of $\cT_p$, which are
presented in Table~\ref{tab:pushing}.

A single random sample is defined as follows. Given $\cT_p$, we add
a face tree at random at each face of $\cT_p$, using an appropriately rescaled version of the
SageMath function \texttt{random()}. We then also add an edge tree to each
edge at random. In doing so, many leaves are added to the original rays of
$\doc$, and the complete set of all rays together with the associated cone
structure define an F-theory base $B$.

This process may be iterated to generate many random samples, and we studied
over $10,000,000$ random samples in this paper.

\subsection{Gauge Group Rank}

In this section we study whether machine learning can accurately predict the rank of the geometric
gauge group in the large ensemble of F-theory geometries.
We will see that it naturally leads to a sharp conjecture for the
gauge group rank. While a version of the conjecture was already proven
in~\cite{Halverson:2017ffz}, exemplifying the process leading to the
conjecture will be important for guiding the genesis of a new conjecture and theorem
in Section~\ref{sec:e6}.

Let $H_i$ be the number of height $i$ leaves in $B$. We seek to train a model $A$ to predict the rank of the resulting gauge group $rk(G)$ on the base $B$
\begin{equation}
B \longrightarrow (H_1,H_2,H_3,H_4,H_5,H_6) \xrightarrow{A} rk(G)\, .
\end{equation}
We perform a $10$-fold cross validation with sample size $1000$ and algorithms LR, LIR, LDA, KNN, CART, NB, SVM. The linear regression gave the best results, having a MAPE of $0.013$. The decision function is
\begin{align}
  rk(G) = 302.54 &-1.1102\times 10^{-16}\, H_1 + 3.9996 \, H_2 + 1.9989\, H_3 \\ \nonumber
  &+ 1.0007\, H_4 + 1.3601\times 10^{-3} \, H_5 + 1.1761\times 10^{-3} \, H_6.
\end{align}
Since height $1$ leaves are facet interior points that are always present, $H_1=38$ for all
samples. This is an important observation, since the lack of variance in $H_1$ means that
the coefficient of $H_1$ can be included in the definition of the intercept by the linear regression. Noting $304=38\times 8$, one can rewrite the above equation equivalently as 
\begin{align}
  rk(G) = -1.46 &+ 8\, H_1 + 3.9996 \, H_2 + 1.9989\, H_3 \\ \nonumber
  &+ 1.0007\, H_4 + 1.3601\times 10^{-3} \, H_5 + 1.1761\times 10^{-3} \, H_6.
\end{align}
This is the output of the linear regression.

We wish to turn the output of machine learning into a sharp conjecture given the rest of the
information that we know about the problem. Our choice to redefine the intercept and the effectively constant $H_1$ term already reflected some underlying knowledge of the F-theory context, as we shall soon see. To go further, note that any leaf can only contribute an integer to the rank of the gauge group; we therefore round all coefficients to the nearest integer. Second, all divisors in $B$ in our language are given by a leaf with some height. Since
seven-branes are wrapped on divisors, it is natural to expect that
with appropriate variables that capture important properties of
divisors, such as $H_i$, the intercept should be close to zero; it is
$-1.46$, which is quite close given that typical values of $rk(G)$ are
$O(2000)$. With these considerations taken into account,
\begin{align}
  rk(G) \simeq  8 \,H_1 + 4\, H_2 + 2\, H_3+ \, H_4.
\end{align}
Recalling that each leaf can only have a geometric gauge group contained in the
set
\begin{equation}
G \in \{E_8, E_7, E_6,F_4, D_4, B_3, G_2, A_2, A_1 \}, 
\end{equation}
it is natural to make the following conjecture based on this analysis:
\vspace{.4cm}
\begin{addmargin}[2em]{2em}
\textbf{Conjecture:} with high probability, height $1$ leaves have gauge group $E_8$, height $2$ leaves have gauge group $F_4$, height $3$ leaves have
  gauge group $G_2$ or $A_2$, and height $2$ leaves have gauge group $A_1$.
\end{addmargin}
\vspace{.4cm}
\noindent This conjecture is the natural output of the discussed
machine learning analysis, linear regression fit, and basic knowledge
of the dataset. The natural questions to ask are what constitutes high
probability, are there important counterexamples that lead to
sub-cases of the conjecture, and how does one go about proving it.

There is one item of critical importance: this conjecture arose out of a set of
random samples, and therefore proving the conjecture may depend on particular properties $P_i$
of the random samples. In this context, ``with high probability'' could mean that
the conjecture depends critically on properties with high probability $P(P_i)$. A natural
proof method, then, is to identify those high probability properties, and try to use them
to prove the conjecture.

By studying random samples, it quickly becomes clear that nearly all $h=1$ leaves carry $E_8$,
and in particular in all random samples computed to date there are $\geq 36$ out of a possible
$38$ $h=1$ leaves that carry an $E_8$. This means that nearly all leaves are built ``above''
$E_8$ roots, meaning that the associated $v$ are linear combinations of $v_i$'s that carry
$E_8$. Since $E_8$ on $h=1$ leaves is so probable, this leads to
\vspace{.4cm}
\begin{addmargin}[2em]{2em}
  \textbf{Refined Conjecture:} Let $v$ be a leaf $v=av_1+bv_2+cv_3$ built on roots $v_{1,2,3}$ whose associated divisors carry $E_8$. Then if the leaf has height $h_v=2,3,4$ its associated
  gauge groups are $F_4$, $\in \{G_2,A_2\}$, and $A_1$, respectively.
\end{addmargin}
\vspace{.4cm}
With this level of refinement, it is possible to do precise calculations leading to
the proof of the conjecture, and a final refinement of the gauge group for height
$3$ leaves.
In fact, a version of this conjecture has already been proven in \cite{Halverson:2017ffz}. The precise result is
\vspace{.4cm}
\begin{addmargin}[2em]{2em}
  \textbf{Theorem:} Let $v$ be a leaf $v=av_1 + bv_2 + cv_3$ with $v_i$ simplex vertices in $F$. If the associated divisors $D_{1,2,3}$
carry a non-Higgsable $E_8$ seven-brane, and if $v$ has height
 $h_v=1,2,3,4,5,6$ it also has Kodaira fiber $F_v=II^*,IV^*_{ns},I^*_{0,ns},IV_{ns},II,-$
and gauge group $G_v=E_8,F_4,G_2,SU(2),-,-$, respectively.
\end{addmargin}
\vspace{.4cm}
The proof is short and is presented in the appendix of \cite{Halverson:2017ffz}.

At this point the reader is probably wondering why we went through a non-trivial exercise to
lead to the formulation of a conjecture, when a version of that conjecture has already been
proven. It is because in this simple result we see a back and forth process using machine
learning and knowledge of the data that led to the formulation of the conjecture, and we
believe this process is likely to be of broader use. That process is:
\begin{enumerate}
\item \emph{\textbf{Variable Selection.}} Based on knowledge of the data, choose input variables $X_i$
  that are likely to determine some desired output variable $Y$. In the example, this
  was recognizing that $X_i=H_i$ may correlate strongly with gauge group.
\item \emph{\textbf{Machine Learning.}} Via machine learning, train a model to predict $Y$ given $X_i$ with high probability. In this example,
  a $10$-fold cross validation was performed, and it was noted that the highest accuracy
  came from a linear regression.
\item \emph{\textbf{Conjecture Formulation.}} Based on how the decision function uses $X_i$ to determine $Y$, formulate a first version of the
  conjecture. In this example, the first version of the conjecture arose
  naturally from the linear regression and basic dataset knowledge.
\item \emph{\textbf{Conjecture Refinement.}} The original conjecture arose from a model
  that was trained on a dataset that is subject to sampling assumptions. Those assumptions
  may lead to high probability properties critical to proving the conjecture; refine accordingly
  based on them. In the example, we used the high frequency of $E_8$ on the ground.
\item \emph{\textbf{Proof.}} After iterating enough times that the conjecture is precise and
  natural calculations or proof steps are obvious, attempt to prove the conjecture.
\end{enumerate}
We will use this procedure to produce new results in Section~\ref{sec:e6}.

\section{Conjecture Generation: $E_6$ Sectors in F-theory Ensembles\label{sec:e6}}

Recently, an ensemble of $\frac43 \times 2.96 \times 10^{755}$
F-theory geometries was presented and universality properties
regarding the gauge group were derived. There, it was shown that
certain geometric assumptions lead to the existence of a gauge group
$G$ of rank $rk(G)\geq 160$ with certain detailed properties that
correlated with the heights of certain ``leaves'' that divisors in an
algebraic threefold. The simple factors in the generically semi-simple
group $G$ were $G_i\in \{E_8,F_4,G_2,A_1\}$, which interestingly only
have self-conjugate representations.  However, $E_6$ and $SU(3)$ may
also exist in this ensemble. In certain random samples they exist with
probability $\simeq 1/1000$, but the conditions under which $E_6$ or
$SU(3)$ existed were not identified at that time.
In this section we wish to use supervised learning to study the
conditions under which $E_6$ exists in the ensemble. We will name
subsections in this section according to the general process
outlined in Section~\ref{sec:rk}.

\subsection{Variable Selection\label{subsec:machinevars}}

The results of~\cite{Halverson:2017ffz} demonstrate that the gauge
group on low lying leaves depends on the heights of trees placed at
various positions around the polytope. 
Yet for the $E_6$ problem that we study, training a model on tree heights did not give as
accurate results as we had expected, and led us to consider other
natural variables on which to train the model. This is the process of
variable selection.

The problem at hand, and for which we are developing new variables, is
the problem of determining whether a particular leaf has a particular
gauge group on it. This question has a yes or no answer, and therefore
this is a classification problem. We will focus on leaves on the
ground, but with some modifications the basic idea extends to
other leaves as well.

Let $v_1$ be a leaf on the ground. Since $mult_v(f)<4$ or
$mult_v(g)<6$ for consistency of the compactification, the monomials
in $f$ and $g$ with minimal multiplicity along $v$ in $f$ and $g$
satisfy these multiplicity bounds. These must necessarily come from
associated points $m_f\in \Delta_f$ and $m_g\in \Delta_g$ satisfying
$-4\leq m_f\cdot v_1<0$ or $-6\leq m_g\cdot v_1 < 0$.  Therefore,
monomials $m$ with $m\cdot v < 0$ play a central role in determining
the gauge group. On the other hand, the above bounds on $m_f$ and
$m_g$ are necessary for playing a role in determining the gauge group
on $v_1$, but they are not sufficient, since if $m_f\cdot v < -4$
($m_g \cdot v < -6$) for some other $v$ then $m_f\not\in \Delta_f$
($m_g\notin\Delta_g$), and then it does not play any role in
determining the gauge group. This follows from the definition
\eqref{eqn:dfdg} of $\Delta_f$ and $\Delta_g$, since all rays $v$
associated to $B$ appear, not just $v_1$, i.e. the upper half planes
associated with all $v$ may in principle chop out $m_f$ and $m_g$, not
just the upper half plane associated to $v_1$.

The task is to determine those $v$ that could cause $m_f$ or $m_g$
to be chopped off of the original $\Delta_f$ or $\Delta_g$, since the gauge group depends on whether or not
this occurs to the set of monomials with minimal order of vanishing
in $f$ or $g$. The other fact we have at our disposal is that at
least one of the inequalities $m_f \cdot v_1 < 0$
or $m_g \cdot v_2 < 0$ must be satisfied. Let us use the $m_f$
inequality, knowing that similar comments hold for the $m_g$ inequality.
Then, for any $v$ that is a leaf in a tree above $v_1$, we have
$v = a v_1 + b v_2 + c v_3$ and therefore $v$ chops off $m_f$ if
\begin{equation}
m_f \cdot v = a \,m_f \cdot v_1 + b\, m_f \cdot v_2 + c\, m_f \cdot v_3 < -4.
\end{equation}
Since $m_f \cdot v_1 < 0$, it is easy to see that the larger $a$ is
the more likely it is that the latter two terms do not compensate
to satisfy the inequality, in which case $m_f$ is chopped off. There may
be many such leaves built on $v_1$ in this way, and they could
occur above different simplices, i.e. with $v_1$ fixed but with
different $v_2$ and $v_3$. Since whether or not $m_f \in \Delta_f$
depends strongly on leaves $v_i$ with associated $a_i$ built above
$v_1$, the gauge group may also.

With this motivating discussion, let us define the variables on which
we train our models. If $v=av_1 + b v_2 + cv_3$, then we define $a$,$b$,$c$
to be the \emph{height above} $v_1,v_2,v_3$, respectively, which sit on the ground. Let
\begin{equation}
S_{a,v_1} := \{v \in V | v = av_1 + b v_2 + c v_3, \,\,\,a,b,c \geq 0 \} 
\end{equation}
for some fixed value of $a$. This set is easily computed for any
$B\in S_{\Delta^\circ}$, i.e. for any $B$ in the ensemble of \cite{Halverson:2017ffz}. For some $a_{max}$, $S_{>a_{max},v_1}$ is
empty, and therefore elements of $S_{a_{max},v_1}$ are the most likely to
  chop out an $m_f$ or $m_g$ relative to elements of $S_{a< a_{max},v_1}$.
  Furthermore, if the cardinality of $S_{a_{max},v_1}$ is large, there
  are more chances to chop off $m_f$. 
To any $v\in V$ satisfying $v\in \Delta^\circ$ (this latter assumption
is for simplicity), we therefore have a map
\begin{equation}
v \mapsto (a_{max}, |S_{a_{max},v}|) \qquad \qquad \forall v \in \Delta^\circ,
\end{equation}
and these will be the set of training variables that we will use.

In summary, instead of training on the \emph{heights} of leaves distributed
throughout $\Delta^\circ$, we will instead train on the number
of leaves of maximal \emph{height above} each $v \in \Delta^\circ$.
An analytical argument was presented above for why these might be
relevant for the gauge group, but due to the complexity of the sets
$S_{a_{max},v}$ it is not obvious how to concretely extract gauge
group information. This is therefore an ideal test case to employ
machine learning techniques.

\subsection{Machine Learning}

Let us now turn specifically to the study of $E_6$ gauge groups that arise in $S_{\Delta_1^\circ}$. In a million random samples, $E_6$ only
arose on a particular distinguished vertex
\begin{equation}
v_{E_6} = (1,-1,-1),
\end{equation}
which is the only vertex of $\Delta_1^\circ$ not in its biggest facet.
However, $E_6$ arose on this vertex with probability $\sim1/1000$,
and (due in part to potential phenomenological relevance) it is of
interest to understand this result better. Specifically, one
would like to have a better understanding of the conditions under which
$E_6$ arises, and whether the probability is a general result or something
that is specific to assumptions of the random sampling. 

Specifically, we will use these variables to train a model to predict whether or not a given $B$ in a random sample
has $E_6$ on $v_{E_6}$. This model defines a map $A$
\begin{equation}
\Delta_1^\circ \longrightarrow (a_{max},|S_{a_{max,v}}|) \quad \forall v\in \Delta_1^\circ \qquad \xrightarrow{A} \qquad \text{$E_6$ on $v_{E_6}$ or not},
\end{equation}
and the goal is to obtain maximal accuracy. For classification problems,
such as this one, accuracy simply means whether $A$ makes
the correct prediction amongst a class of possibilities, which
in this case is a binary class.

We will train on $20,000$ samples, but in this particular case it is important to change the sampling
prescription slightly. Under a purely random sample, approximately $20$ of the cases sampled would indeed have $E_6$ on $v_{E6}$ and the rest would not. In this
case, the trained algorithm will naturally lead to a constant prediction of no $E_6$
for all inputs, and it would have accuracy $.999$. This is clearly a sub-optimal outcome with a misleading accuracy value. To correct for this possibility we will use random sampling to generate $10,000$ samples with $E_6$ and $10,000$ without, training on the combined set.

\begin{figure}[t]
  \centering
  \includegraphics[scale=.5]{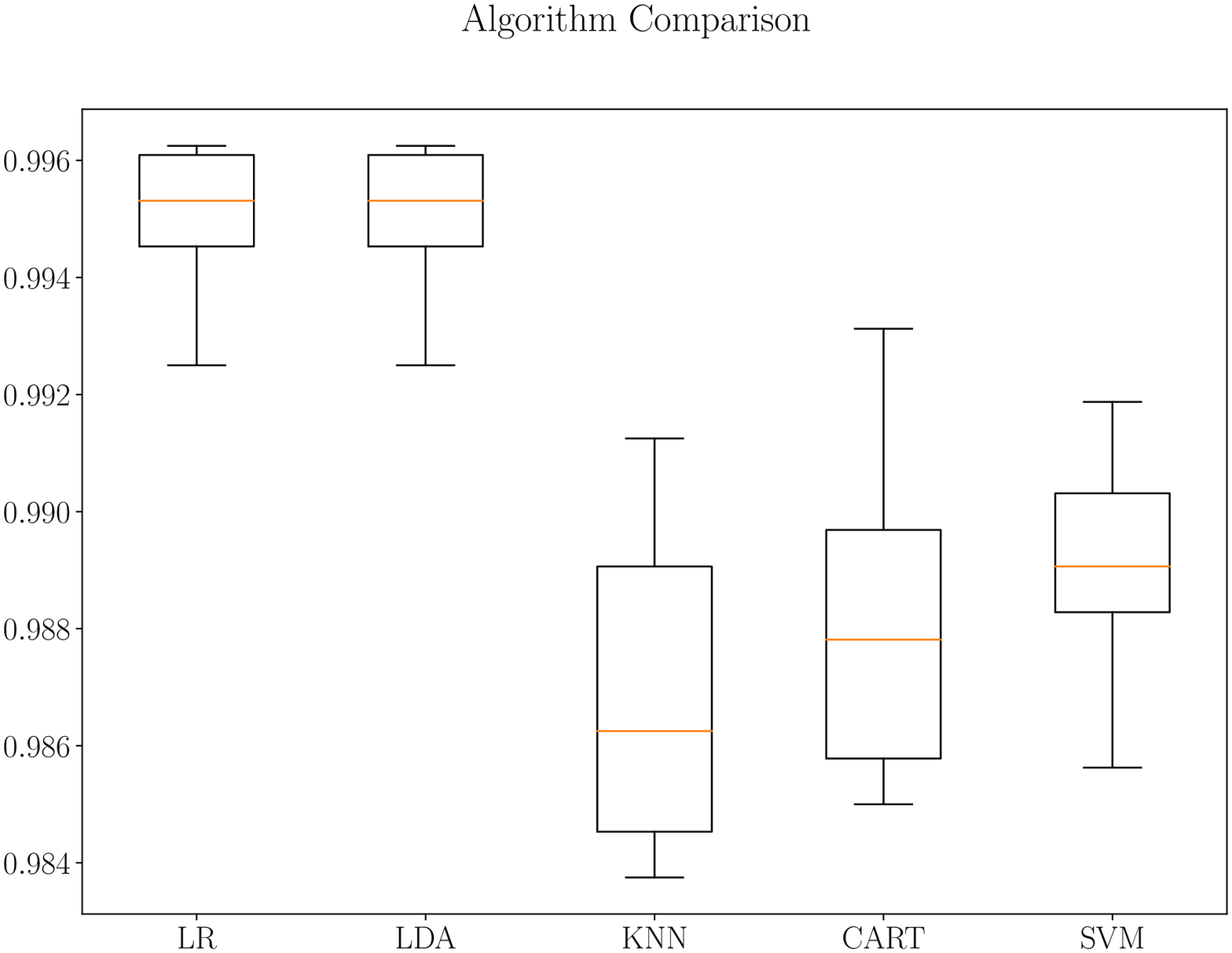}
  \caption{Model comparison from $10$-fold cross validation for $E_6$ sectors. All algorithms perform quite well, though logistic regression
  and linear discriminant analyses give the best results.}
  \label{fig:e6algwhisker}
\end{figure}

Using this dataset, we perform a $10$-fold cross validation using
algorithms LR, LDA, KNN, CART, SVM, and a validation size of $0.2$,
which reduces the training set to have $16,000$ samples and the validation
set to have $4000$.
The results are presented in Figure~\ref{fig:e6algwhisker}, and we
see that all of the models achieve high accuracy, with maximal
median accuracy of $.995$ achieve by the LR and LDA models.

Though we see high accuracy on the enriched samples with $50\%$ $E_6$
on $v_{E_6}$ and $50\%$ not, it is also interesting to ask whether the
models trained on the $50/50$ set can make predictions on an
unenriched set with $E_6$ occurring naturally
with probability $1\simeq 1000$. We trained the models on the
$50/50$ training set, and scored them on the $50/50$ validation
set of size $4000$ and unenriched set of size $20,000$, with
accuracy:
\begin{center}
  \begin{tabular}{cccccc}
    & LR & LDA & KNN & CART & SVM \\ \hline
    $50/50$ Validation Set & $.994$ & $.994$ & $.982$ & $.987$ & $.989$ \\
    Unenriched Set & $.988$ & $.988$ & $.981$ & $.988$ & $.983$. \\
  \end{tabular}
\end{center}
We see that all models accurately predict whether or not there is
$E_6$ on $v_{E6}$ in the unenriched data, even though the models were
trained on the $50/50$ enriched data.  We have done this
using $(a_{max},|S_{max,v}|)$ for all $v$ in $\Delta_1^\circ$, which is
a set of $2\times 38 = 76$ integers.

It is natural, however, to ask whether the result is ultimately
controlled by some subset of these integers. As discussed in Section
\ref{sec:mlforst}, analyses of this sort are known as
\emph{dimensionality reduction}. Applied in this case, a particular
type of dimensionality reduction known as a factor analysis
demonstrates that, to high accuracy, the question of whether or not $E_6$ is on
$v_{E_6}$ is determined by $(a_{max},|S_{a_{max},v_{E6}}|)$. Specifically, the
factor analysis identified that a particular linear combination of the original
training variables, which were$(a_{max},|S_{a_{max},v}|) \, \, \forall v \in \dc$,
determined whether or not there was an $E_6$ on $v_{E_6}$. That linear
combination only had non-negligible components along the pair of
integers $(a_{max},|S_{a_{max},v_{E6}}|)$.
It is natural to
expect this, given our previous discussion that motivated the use of
these variables in the first place; nevertheless, the factor analysis
underscores the relevance of this feature.

Restricting the training data to $(a_{max},|S_{a_{max},v_{E6}}|)$, we performed
  identical analyses to the previous ones with this restricted set of inputs
  per example, and find even better accuracy:
\begin{center}
  \begin{tabular}{cccccc}
    & LR & LDA & KNN & CART & SVM \\ \hline
    $50/50$ Validation Set & $.994$ & $.994$ & $.994$ & $.994$ & $.994$ \\
    Unenriched Set & $.988$ & $.988$ & $.988$ & $.988$ & $.983$. \\
  \end{tabular}
\end{center}
Thus, we see that there is a single pair of variables, $(a_{max},|S_{max,v_{E6}}|)$,
that determines to high accuracy whether or not there is an $E_6$ factor
on $v_{E6}$! Machine learning has validated the loose ideas as to why
$(a_{max},|S_{max,v}|)$ might in some cases be relevant variables from which
to predict gauge groups.

\subsection{Conjecture Formulation}
Heartened by the accuracy of the model and the simplicity of the input
data, we would like to use it to formulate a conjecture that can be proven rigorously.

To formulate a conjecture using the machine learned model there are
two natural paths. One is to look under the hood of the
algorithm used to fit the model, study the decision function, and see
how it makes predictions given certain inputs. This was the path taken in Section~\ref{sec:rk}. For datasets with low dimensionality (or low effective dimensionality after dimensionality reduction) it may be possible to directly examine the predictions for each input and see if there is any obvious trend. For the dimensionally reduced input data that we just discussed, there are in fact only $32$ unique pairs $(a_{max},|S_{a_{max},v_{E6}}|)$ in the $20,000$ samples, suggesting that human input may be feasible at this step in the conjecture-generating process.

\begin{table}[th]
  \centering
  \begin{tabular}{|cccc|}
    \hline
    $a_{max}$ & $|S_{a_{max},v_{E6}}|$ & Pred. for $E_6$ on $v_{E_6}$ & Hyperplane Distance\\ \hline
$4$&$5$&No&$0.88$ \\
$4$&$6$&No&$0.29$ \\
$4$&$7$&Yes&$-0.31$ \\
$4$&$8$&Yes&$-0.90$ \\
$4$&$9$&Yes&$-1.50$ \\
$4$&$10$&Yes&$-2.09$ \\
$4$&$11$&Yes&$-2.69$ \\
$4$&$12$&Yes&$-3.28$ \\
$4$&$13$&Yes&$-3.88$ \\
$4$&$14$&Yes&$-4.47$ \\
$4$&$15$&Yes&$-5.07$ \\
$4$&$16$&Yes&$-5.67$ \\
$4$&$17$&Yes&$-6.26$ \\
$4$&$18$&Yes&$-6.85$ \\
$4$&$19$&Yes&$-7.45$ \\
$4$&$20$&Yes&$-8.04$ \\
$4$&$21$&Yes&$-8.64$ \\
$4$&$22$&Yes&$-9.23$ \\
$4$&$23$&Yes&$-9.83$ \\
$4$&$24$&Yes&$-10.42$ \\
$5$&$1$&No&$7.34$ \\
$5$&$2$&No&$6.75$ \\
$5$&$3$&No&$6.15$ \\
$5$&$4$&No&$5.56$ \\
$5$&$5$&No&$4.96$ \\
$5$&$6$&No&$4.37$ \\
$5$&$7$&No&$3.78$ \\
$5$&$8$&No&$3.18$ \\
$5$&$9$&No&$2.59$ \\
$5$&$10$&No&$1.99$ \\
$5$&$11$&No&$1.40$ \\
$5$&$12$&No&$0.80$ \\ \hline
  \end{tabular}
  \caption{Predictions of our logistic regression model as a function
  of $(a_{max},|S_{max,v_{E_6}}|)$.}
  \label{tab:e6predandhypdist}
\end{table}

Utilizing a logistic regression to train a model on these $20,000$ samples,
the model makes predictions for whether or not there is an $E_6$ on $v_{E_6}$
as a function of $(a_{max},|S_{a_{max},v_{E6}}|)$, with the predictions given in
Table \ref{tab:e6predandhypdist}. There is an obvious trend: it always
predicts no for $a_{max}=5$, and usually predicts no for $a_{max}=4$. This
is highly suggestive that whether $a_{max}$ is $4$ or~$5$ for $v_{E6}$
correlates strongly with whether or not there is an $E_6$. The hyperplane distance
is an intrinsic measure of the confidence of the prediction based on the
logistic regression.
The conclusions of this analysis lead to
\vspace{.4cm}
\begin{addmargin}[2em]{2em}
  \textbf{Conjecture:} If $a_{max}=5$ for $v_{E6}$, then $v_{E6}$ does not
  carry $E_6$. If $a_{max}=4$ for $v_{E6}$ it may or may not carry $E_6$,
  though it is more likely that it does.
\end{addmargin}
\vspace{.4cm}
The initial conjecture is rough, as was the initial conjecture of Section~\ref{sec:rk}.

\subsection{Conjecture Refinement and Proof}
We now attempt a conjecture refinement based on the sampling assumptions, the
high probability properties to which they lead, and general knowledge
of the problem at hand.

As discussed, e.g., in~\cite{Anderson:2014gla}, a necessary
condition for $v_{E6}$ to carry $E_6$ is that
\begin{equation}
g = x_{E_6}^4(m^2 + \dots),
\end{equation}
where $x_{E_6}$ is the homogeneous coordinate associated to $v_{E_6}$
and $g_4=m^2$ is a single monomial if $B$ is toric. This single monomial corresponds to a single
$\tilde m\in\bZ^3$ satisfying
\begin{equation}
\tilde m \cdot v_{E_6} + 6 = 4.
\end{equation}
In $40,000$ random samples, whenever $E_6$ arose the example had
\begin{equation}
\tilde m = (-2,0,0).
\end{equation}
Henceforth, by $\tilde m$ we will mean precisely this vector in
$\bZ^3$. Finally, in random samples we have empirically only seen the gauge group
$G$ on $v_{E_6}$ arising as $G\in \{E_6, E_7, E_8\}$, suggesting that
$E_6$ only arises with high probability if $\tilde m = (-2,0,0)$. With
some hard work, this probability could be computed, but we leave this
for future work and instead take it as a hypothesis for our refined
conjectures.

These valuable pieces of information, together with our model analysis,
suggests that $\tilde m$ is critical to obtaining $E_6$, and furthermore that its
should correlate strongly with $a_{max}$ being $4$ or $5$. This leads to a
refined conjecture based on evidence from the samples:
\vspace{.4cm}
\begin{addmargin}[2em]{2em}
  \textbf{Refined Conjecture:} Suppose that with high probability the group $G$
  on $v_{E_6}$ is $G\in \{E_6,E_7,E_8\}$ and that $E_6$ may only arise with  $\tilde m = (-2,0,0)$. Then
  there are two cases related to determining $G$.
  \begin{itemize}
    \item[a)]  If $a_{max}=5$, $\tilde m$
      cannot exist in $\Delta_g$ and the group on $v_{E6}$ is above $E_6$.
  \item[b)] If $a_{max}=4$, $\tilde m$ sometimes exists in $\Delta_g$. If it does then there is an $E_6$ on $v_{E_6}$, and if it does not there is an $E_7$ or $E_8$ on $v_{E_6}$.
    
    \end{itemize}
  \end{addmargin}
  Attempting to prove this quickly leads to additional realizations that give a final
  conjecture:
\vspace{.1cm}
\begin{addmargin}[2em]{2em}
  \textbf{Theorem:} Suppose that with high probability the group $G$
  on $v_{E_6}$ is $G\in \{E_6,E_7,E_8\}$ and that $E_6$ may only arise with  $\tilde m = (-2,0,0)$.  Given these assumptions, there are three cases that determine whether
  or not $G$ is $E_6$.
  \begin{itemize}
    \item[a)]  If $a_{max}\geq 5$, $\tilde m$
      cannot exist in $\Delta_g$ and the group on $v_{E_6}$ is above $E_6$.
    \item[b)] Consider $a_{max}=4$. Let $v_i= a_i v_{E_6}+b_iv_2 + c_i v_3$ be
      a leaf built above $v_{E_6}$, and $B=\tilde m \cdot v_2$ and $C=\tilde m \cdot v_3$.
      Then $G$ is $E_6$ if and only if $(B,b_i)>0$ or $(C,c_i)>0$ $\forall i$.
      Depending on the case, $G$ may or may not be $E_6$.
    \item[c)] If $a_{max}\leq 3$, $\tilde m \in \Delta_g$ and the group is $E_6$.
    \end{itemize}
  \end{addmargin}
\begin{addmargin}[2em]{2em}
  \textbf{Proof:} We will proceed by a number of direct computations
  to determine the relationship between $a_{max}$ and whether
  $\tilde m \in \Delta_g.$ Recall that
  $\tilde m \in \Delta_g \leftrightarrow \tilde m\cdot \tilde v_i + 6
  \geq 0 \,\, \forall \tilde v_i$, where the $\tilde v_i$ are any
  leaves. Direct computation shows that
  $\tilde m\cdot \tilde v_i \in \{-2,0,2\}$ for those
  $\tilde v_i\in \Delta_1^\circ$, and that $v_{E_6}$ is the only
  interior point of $\Delta_1^\circ$ satisfying
  $\tilde m \cdot v_{E_6}=-2$. Therefore, any leaf $v$ that cuts
  $\tilde m$ out of $\Delta_g$, i.e.  $\tilde m \cdot v + 6 < 0$,
  necessarily has a component along $v_{E_6}$. Let
  $v=av_{E_6} + b v_2 + c v_3$, with $a,b,c\geq 0$; normally we
  require strict inequality, but do not here so that $v$ may be a leaf
  in a face tree or and edge tree. Given the above set $\{-2,0,2\}$,
  this yields $-2a+6\leq \tilde m \cdot v + 6 \leq -2a + 2(b+c) + 6$.
  We study cases of this general inequality. If $a_{max}=5$ there is
  at least one leaf with $a=5$, and our bound $a+b+c\leq 6$ implies
  $b+c=1$.  Then $\tilde m \cdot v + 6 \leq -10 + 2 + 6 < 0$ and
  therefore $\tilde m \notin\Delta_g$.  This enhances the gauge group
  on $v_{E_6}$ beyond $E_6$, proving $a)$. On the other hand, if
  $a_{max}=3$, $\tilde m \cdot v + 6 \geq -6 + 6 = 0$ and
  $\tilde m \in \Delta_g$, proving c).  The case that
  requires some work is b), which has $a_{max}=4$. Let $B, C$ be
  $\tilde m \cdot v_2$, $\tilde m \cdot v_3$. The most constraining
  leaves are those $v$ with $a=a_{max}$, in which case
  $\tilde m \cdot v + 6= -2 + b B + c C$. From above, we have
  $B,C\in \{0,2\}$ and $(b,c)\in \{(1,0),(0,1),(1,1)\}$. Then
  $\tilde m \cdot v + 6 \geq 0 \leftrightarrow$ $b$ and $B$ are
  non-zero or $c$ and $C$ are non-zero. This must occur for all leaves
  $v$, in which case $G$ is $E_6$, proving b).
\end{addmargin}
\vspace{.4cm}
This theorem is a stronger, rigorous version of the basic result from the model we
trained with machine learning, namely that if $a_{max}=5$ then the gauge group $G$ on it is above $E_6$, whereas it may or may not be $E_6$ if $a_{max}=4$.

It is interesting that this result does not depend on triangulation,
instead only that a random
sampling on some triangulation give $G\in \{E_6,E_7,E_8\}$ with high probability and that $E_6$
arise with $\tilde m = (-2,0,0)$. If these assumptions hold in any particular triangulation,
then the likelihood of $a)$, $b)$, or $c)$ occurring can be computed explicitly based on
the detailed cone structure. Any three-dimensional cone containing $v_{E_6}$ is determined
by a $3\times 3$ matrix $M=(v_{E_6},v_2, v_3)$ subject to the constraint $|det(M)|=1$, and
from this data $B$ and $C$ can be determined. Without loss of generality
we can choose $B \geq C$, and directly compute $(B,C) \in \{(2,2),(2,0),(0,0)\}$.
Note that since a leaf in a face tree with height $4$ above $E_6$ is $v=4v_{E_6}+bv_1+cv_2$
and has $b,c>0$, such a leaf can cut out $\tilde m$ only in the case $(B,C)=(0,0)$.
This result is triangulation independent, and we leave the study of other triangulations
to future work.

We would like to study the conditions of the theorem in the triangulation from which we
built our random samples. The three cones in this triangulation that contain $v_{E_6}$ are presented in Table \ref{tab:ve6pushing}
\begin{table}
  \centering
  \begin{tabular}{|ccc|}

    \hline $v_{E_6}$ & $v_2$ & $v_3$ \\ \hline
$(1, -1, -1)$&$(-1, -1, 0)$&$(0, -1, -1)$\\
$(1, -1, -1)$&$(-1, 0, -1)$&$(0, -1, -1)$\\
$(1, -1, -1)$&$(-1, 0, -1)$&$(0, 0, -1)$\\
$(1, -1, -1)$&$(-1, -1, 0)$&$(0, -1, 0)$\\
$(1, -1, -1)$&$(-1, 2, -1)$&$(0, 0, -1)$\\
$(1, -1, -1)$&$(-1, 2, -1)$&$(0, 1, -1)$\\
$(1, -1, -1)$&$(-1, 4, -1)$&$(0, 1, -1)$\\
$(1, -1, -1)$&$(-1, 4, -1)$&$(0, 2, -1)$\\
$(1, -1, -1)$&$(-1, -1, 2)$&$(0, -1, 0)$\\
$(1, -1, -1)$&$(-1, -1, 2)$&$(0, -1, 1)$\\
$(1, -1, -1)$&$(-1, -1, 4)$&$(0, -1, 1)$\\
$(1, -1, -1)$&$(-1, -1, 4)$&$(0, -1, 2)$\\
$(1, -1, -1)$&$(-1, 0, 4)$&$(0, -1, 2)$\\
$(1, -1, -1)$&$(-1, 0, 4)$&$(0, 0, 1)$\\
$(1, -1, -1)$&$(-1, 2, 2)$&$(0, 0, 1)$\\
$(1, -1, -1)$&$(-1, 2, 2)$&$(0, 1, 0)$\\
$(1, -1, -1)$&$(-1, 4, 0)$&$(0, 1, 0)$\\
    $(1, -1, -1)$&$(-1, 4, 0)$&$(0, 2, -1)$\\ \hline
  \end{tabular}
  \caption{The three-dimensional cones of the pushing triangulation $\cT_p$ of
    $\doc$ that contain $v_{E_6}$.}
  \label{tab:ve6pushing}
\end{table}
and all are of $(B,C)=(2,0)$ type. Therefore in this triangulation
face leaves with $a_{max}=4$ cannot cut out $\tilde m$. Let us
consider edge leaves, $v=av_{E_6}+b v_2$, which we will refer to as an
$(a,b)$ edge leaf. Leaves above edges with $a_{max}=4$ have
$v=4v_{E_6}+b v_2$ may be able to cut out $\tilde m$ from $\Delta_g$. From part $b)$ of the theorem, this
occurs when $B=\tilde m \cdot v_2=0$, and the only possibility for $b$
is $b=1$. There are $18$ two dimensional cones containing $v_{E_6}$ in our
ensemble, $9$ with $B=2$ and $9$ with $B=0$.

This theorem and ensuing discussion imply that $E_6$ exists
on $v_{E_6}$ in this triangulation if and only if there are no $(5,1)$
edge leaves and there are no $(4,1)$ edge leaves on edges with
$B=0$. A pertinent fact is that edge trees with $(5,1)$ edge leaves always also have
$(4,1)$ edge leaves, and therefore the stated condition occurs if and only if there
are no $(5,1)$ edge leaves on $B=2$ edges and no $(4,1)$ edge leaves on $B=0$ edges.
Of the $82$ possible edge trees, $36$ have $(4,1)$ leaves, and $18$ have $(5,1)$ leaves.
The probability of $E_6$ on this triangulation $T$ of $\Delta_1^\circ$ should then be
\begin{equation}
P(E_{6} \,\text{on}\, v_{E_6}\, \text{in}\, T) = \left(1-\frac{36}{82}\right)^9\left(1-\frac{18}{82}\right)^9\simeq .00059128.
\end{equation}
This prediction should be checked against random samples. Performing five
independent sets of two million
random samples each, the predicted number of models with $E_6$ using this theorem and
associated probability,
compared to the results from random samples, is
\begin{align}
\text{From Theorem}&: \qquad.00059128 \times 2 \times 10^6=1182.56 \nonumber \\
\text{From Random Samples}&: \qquad 1183,\,1181,\,1194,\,1125,\,1195. 
\end{align}
Some statistical variance is naturally expected when sampling, but the
agreement is exceptional. Since the probability is computed from a
theorem and is reliable when comparing to a random sample, we compute
the number of models with $E_6$ on $v_{E_6}$ given this triangulation:
\begin{equation}
  \text{Number of $E_6$ Models on $T$} = .00059128\times \frac 13 \times 2.96 \times 10^{755} = 5.83\times 10^{751}.
\end{equation}
It would be interesting to study phenomenological aspects of these models and whether
the probability of $E_6$ changes in different triangulations of $\Delta_1^\circ$. We leave
this to future work.

\section{Conclusions}

  In this paper we have utilized machine learning to study the string landscape.
  We have exemplified two concepts that we believe will be of broad use in understanding
  the landscape: deep data dives and conjecture generation.

  In a deep data dive, a model trained by machine learning on a subset
  of a dataset allows for fast and accurate predictions outside of the
  training set, allowing for fast exploration of the set. In some
  cases this exploration would not be possible without the model. The
  example of Section \ref{sec:3dPoly} is a deep data dive that studies
  triangulations of $3d$ reflexive polytopes. There, we used machine
  learning and 
  $10$-fold cross validation to optimize a model, eventually selecting an optimized
  decision tree for our study. This decision tree
  accurately predicts the average number of fine regular
  star triangulations per polytope at a given value of $\hoo$ of the
  associated toric variety, $\overline{n_{FRST}}(\hoo)$. These results were already known, providing
  a basis for evaluating the machine learning results. We found that the decision tree accurately predicts $\overline{n_{FRST}}(\hoo)$ for five values of $\hoo$
  beyond the training set, though the behavior is erratic at higher $\hoo$ likely
  due to being in the tail of the distribution. However, the extrapolation of reliable machine
  learned data to higher $\hoo$ accurately predicts the known order of
  magnitude $\overline{n_{FRST}}(\hoo=35)\sim 10^{15-16}$. In \cite{Upcoming}
  machine learning will be used to study Calabi-Yau threefolds
  in the Kreuzer-Skarke set.

  In conjecture generation, machine learning is used to extract
  hypotheses regarding data features that can lead to the formulation
  of a sharp conjecture.  We found a common procedure that worked for
  the examples of Section \ref{sec:rk} and \ref{sec:e6}: variable
  selection, machine learning, conjecture formulation, conjecture
  refinement, proof. Each of the elements is described in Section
  \ref{sec:rk}, and the section headings of Section \ref{sec:e6} are chosen
  according to this procedure. In Section \ref{sec:rk} we studied the
  rank of the geometrically non-Higgsable gauge group in an ensemble
  of $\frac43 \times 2.96\times 10^{755}$ F-theory geometries.  We
  used machine learning and $10$-fold cross validation to optimize a
  model, and found that a simple linear regression performed
  best. This naturally led to a conjecture that the rank of the gauge
  group depends critically on the number of leaves of a given height
  in the geometry, and a version of this conjecture had already been
  proven.  In Section \ref{sec:e6} we studied the appearance of an
  $E_6$ factor on a distinguished vertex $v_{E_6}$ in the same
  ensemble. We again utilized machine learning and $10$-fold cross
  validation to optimize a model, finding that a logistic regression
  made the most accurate predictions. This led to the generation of a new
  conjecture regarding when $E_6$ occurs on $v_{E_6}$, which was then proven
  and compared to $10,000,000$ random samples, with good agreement.
  Both of these sections
  demonstrated the utility of machine learning in generating
  conjectures, and underscore the importance of supervision in supervised
  machine learning: variable selection and dataset knowledge were central
  to improving the performance of the machine learned models.

  We find machine learning a promising way to address big data problems in the string
  landscape, and find it particularly encouraging that these numerical techniques
  may lead to rigorous results via conjecture generation.

\acknowledgments{We thank Ross Altman, Tina Eliassi-Rad, Cody Long,
  and Ben Sung for useful discussions.  J.H. is supported by NSF Grant
  PHY-1620526. B.D.N. is supported by NSF Grant PHY-1620575.}

\bibliographystyle{JHEP}
\bibliography{refs}

\end{document}